\documentclass[letterpaper,12pt]{article}
\usepackage[hmargin=1in,vmargin=1in]{geometry}

\usepackage{graphicx} 

\usepackage{listings}
\usepackage{amsmath,amsfonts,amssymb,amsthm,epsfig,xspace,graphics,graphicx,mathtools}

\usepackage{setspace}
\usepackage{framed}
\usepackage{hyperref}
\usepackage{cancel}

\usepackage[round,authoryear]{natbib}
\usepackage[normalem]{ulem}

\graphicspath{{./Figures/}}

\newenvironment{Remark}{
\vspace{10pt}
\vspace{10pt}
\noindent{\bf Remark.}\
}{
  \par \vspace{10pt}
  \vspace{10pt}
}

\newcommand{\trans}{{\mskip-2mu\scriptscriptstyle\top}} 

\usepackage{pdfpages} 

\newcommand{\Def}{\overset{\text{def}}{=}}

\newcommand{\Bt}{\calB_t}  
\newcommand{\Br}{\calB_\mat}  
\newcommand{\St}{\calS_{\breve{\bft}}}  
\newcommand{\Stmat}{\calS_{\breve{\bft}_\mat}}

\newcommand{\Su}{\calS_{\breve{\bfu}}}  
\newcommand{\Sd}{\calS_d}

\newcommand{\nup}{\mbox{$\nu^p$}}

\newcommand{\xiR}{\boldsymbol{\xi}_\mat}

\newcommand{\SxiR}{\calS_{\xiR}} 
\newcommand{\Sxi}{\calS_{\boldsymbol{\xi}}}

\newcommand{\zed}{{\bf 0}}
\newcommand{\id}{{\bf 1}}

\newcommand{\vvs}{\mskip1mu}
\newcommand{\mat}{\text{\tiny R}}%

\newcommand{\grad}{\text{grad}\,}
\newcommand{\divx}{\text{div}\vvs} 

\newcommand{\Grad}{\nabla}

\newcommand{\tendot}{\mskip-3mu:\mskip-2mu}

\newcommand{\calB}{\mathcal{B}}%
\newcommand{\calS}{\mathcal{S}}%
\newcommand{\bfA}{{\bf A}}%
\newcommand{\bfb}{{\bf b}}\newcommand{\bfB}{{\bf B}}%
\newcommand{\bfC}{{\bf C}}%
\newcommand{\bfD}{{\bf D}}%
\newcommand{\bfE}{{\mathbb E}}%
\newcommand{\bfELN}{{\bf E}}%
\newcommand{\bfF}{{\bf F}}%
\newcommand{\bfH}{{\bf H}}%
\newcommand{\bfK}{{\bf K}}%
\newcommand{\bfL}{{\bf L}}%
\newcommand{\bfM}{{\bf M}}%
\newcommand{\bfn}{{\bf n}}\newcommand{\bfN}{{\bf N}}%
\newcommand{\bfR}{{\bf R}}%
\newcommand{\bft}{{\bf t}}\newcommand{\bfT}{{\bf T}}%
\newcommand{\bfu}{{\bf u}}\newcommand{\bfU}{{\bf U}}%
\newcommand{\bfv}{{\bf v}}%
\newcommand{\bfW}{{\bf W}}%
\newcommand{\bfx}{{\bf x}}%
\newcommand{\bfxi}{\boldsymbol{\xi}}%
\newcommand{\bfchi}{\boldsymbol{\chi}}%
\newcommand{\Schi}{\calS_{\boldsymbol{\breve{\chi}}}}%

\newcommand{\skw}{\hbox{\rm skw}\mskip3mu}
\newcommand{\sym}{\hbox{\rm sym}\mskip3mu}
\newcommand{\Tr}{\hbox{\rm tr}\mskip2mu}

\newcommand{\Div}{\hbox{\rm Div}\mskip2mu}  

\renewcommand{\ddot}{\dot{d}}
\newcommand{\zetaR}{\zeta_{\mat}}
\newcommand{\HR}{\mathcal{H}_\mat}
\newcommand{\HS}{\mathcal{H}}

\newcommand{\epsilonR}{{\varepsilon}^{f}_\mat}
\newcommand{\epsilonS}{{\varepsilon}^{f}}

\newcommand{\omegaR}{{\omega}_\mat}

\title{A finite element implementation of a large deformation gradient-damage theory for fracture with Abaqus user material subroutines}

\author{Keven Alkhoury${^\dagger}$\thanks{Corresponding authors: keven$\_$alkhoury@brown.edu \& vikas$\_$srivastava@brown.edu}, Shawn A. Chester${^\ddagger}$, and Vikas Srivastava${^\dagger}\footnotemark[1]$\\
\ \\
${^\dagger}$ School of Engineering\\
Brown University\\
184 Hope Street, Providence, RI 02906 USA\\
\ \\
${^\ddagger}$ Mechanical Engineering\\
New Jersey Institute of Technology\\
Newark, NJ 07102 USA
}

\begin{document}

\maketitle


\begin{abstract}

Recent advancements in computations have enabled the application of various modeling approaches to predict fracture and failure, such as the gradient-damage (phase-field) method.
Several existing studies have leveraged the heat equation solver in Abaqus to model gradient-damage,  due to its mathematical resemblance to the heat equation. Particular care is required when extending the approach to large deformation scenarios due to differences in the \emph{referential} and \emph{spatial} configurations, especially since the heat equation in Abaqus is solved in the spatial configuration, whereas most gradient-damage frameworks are formulated in the referential configuration.
This work provides a pedagogic view of an appropriate  Abaqus implementation of a gradient-damage theory for fracture in materials \emph{undergoing large deformation} using Abaqus UMAT and UMATHT user subroutines. Key benchmark problems from the literature are used to demonstrate the robustness of our implementation across various materials exhibiting different constitutive behaviors, such as non-linear elasticity, linear elasticity, and large deformation rate-dependent plasticity, ensuring its applicability regardless of the specific material constitutive choice. The details of the implementation, along with the codes, which are a direct outcome of this work, are also provided.

\end{abstract}



\section{Introduction}

Failure and fracture have long played a central role in shaping the world around us. From natural evolution to human engineering, the ability of materials and structures to resist breaking when subjected to loadings has been vital to survival, safety, and progress.
Over the last century, rapid technological advancements have deepened our understanding of the mechanics of materials and sparked scientific interest in fracture mechanics, with groundbreaking development tracing back to \citet{griffith1921vi} whose work laid the foundation for the field.

The literature related to phase-field modeling of fracture, originally formulated to include quasi-static brittle settings \citep{francfort1998revisiting,bourdin2000numerical,miehe2010phase,miehe2010thermodynamically}, has seen significant theoretical development including dynamic fracture \citep{borden2012phase}, ductile fracture \citep{ambati2015phase,ambati2016phase}, as well as large deformation fracture, often encountered in polymeric materials \citep{miehe2014phase,talamini2018progressive,narayan2021fracture,konale2025modeling}.

More recently, advancements in computation have enabled the application of various modeling approaches to predict fracture and failure, such as the gradient-damage (phase-field) method,\footnote{While the term ``phase-field" can refer to a broader class of methods depending on the application (i.e., solidification, fracture, multi-phase systems, etc.), we use in this work the terms ``gradient-damage" and ``phase-field" interchangeably to describe the regularized formulation of damage evolution and fracture in solids.} which is capable of modeling damage and failure in a wide range of materials, including metals \citep{miehe2015phase1,chu2019unified}, rubbers \citep{miehe2014phase,talamini2018progressive,loew2019rate}, glass \citep{mehrmashhadi2020validating,egboiyi2022mechanistic}, ceramics \citep{wilson2013phase,li2022three}, fiber-reinforced composites \citep{zhang2019phase,konica2023phase}. It also has the capability of predicting various multiphysics problems with a variety of failure mechanisms, such as hydrogen-embrittlement \citep{kristensen2020phase,dinachandra2022adaptive,cui2022generalised} and thermal degradation \citep{miehe2015phase2,svolos2020thermal,najmeddine2024physics}, to name a few. 

As a reminder, the phase-field method utilizes a scalar variable representing ``damage'' denoted as $d$ and its gradient, $\Grad d$, to regularize the sharp crack topology by introducing a diffusive damage zone within a characteristic length scale $l$. Some key advantages of this approach are resolving numerical challenges associated with sharp cracks and discontinuities and facilitating mesh-independent crack propagation simulations. Consequently, significant effort has been dedicated by the computational mechanics community to implementing the gradient-damage theory in both commercially available programs such as ABAQUS and COMSOL, and open-source programs such as FEniCS \citep[cf. eg.,][and references within]{msekh2015abaqus,liu2016abaqus,molnar20172d,zhou2018phase,fang2019phase,hirshikesh2019fenics,li2020variational,molnar2020open,wu2020comprehensive,kristensen2020phase,najmeddine2024efficient,hai2024dynamic,george2025phase}.

In what follows, we focus our attention on Abaqus due to its widespread popularity and extensive use within the mechanics community \citep{zhong2021higher,navidtehrani2021simple,navidtehrani2021unified,konale2023large,vaishakh2024hygroscopic}. Notably, most existing studies utilizing Abaqus to model phase-field fracture have involved the use of multiple subroutines, often incorporating a user element (UEL) \citep{lee2023finite, konale2025modeling}, which are complex to implement, or are restricted to linear constitutive behavior, i.e., small strain kinematics. The excellent contributions of \citet{navidtehrani2021simple,navidtehrani2021unified} address the first limitation by providing a simple yet robust implementation using only a user material subroutine (UMAT). This eliminates the complexities of UEL programming and enables the use of Abaqus built-in features. However, their approach remains limited to brittle failure under small-strain kinematics and cannot capture the large-deformation fracture and failure behavior critical for predicting the response of many engineering materials.

The objective of this work is to address the shortcomings listed above by providing a pedagogic view of an appropriate Abaqus implementation of a gradient-damage theory for fracture in materials \emph{undergoing large deformation} using Abaqus user material subroutines UMAT and UMATHT.\footnote{As will be clarified shortly, the UMATHT subroutine enables the modification of the thermal constitutive behavior and is essential for capturing the mechanics associated with large deformation kinematics, due to differences in the \emph{referential} and \emph{spatial} configurations.} This allows for accurately solving damage and failure problems without the need for writing complex UELs. We note that the intention of this work is not to develop a new damage framework but rather to implement existing theories efficiently and correctly,  using Abaqus user material subroutines (reviewed and summarized in Section \ref{sec:Summary}). We accomplish this by carefully reviewing the analogy between the heat equation implemented in Abaqus, and the partial differential equation (PDE) governing gradient-damage theory for large deformation fracture in Section \ref{sec:FE_Implementation}. Next, we provide the detailed finite element implementation using the user material subroutines UMAT and UMATHT \citep{AbqStan}, which was verified against the user element implementation provided by \citet{lee2023finite} and applied to benchmark problems from the literature in Section \ref{sec:Applications}, including a penny-shaped specimen in tension, a single edge notched tension (SENT) test, and a single edge notched beam (SENB) test. Lastly, we provide the Abaqus user material subroutines and the corresponding input files as supplemental materials to this paper.

\section{Summary of the gradient-damage theory}\label{sec:Summary}

In this section, we summarize the gradient-damage theory for solids undergoing large deformation following the existing work in literature \citep[][and references within]{miehe2014phase,talamini2018progressive,narayan2021fracture,lee2023finite, konale2025modeling}. For further details related to the theoretical framework, the reader is referred to the original works cited above.

\subsection{Kinematics}

Consider an undeformed body $\calB_\mat$ identified with the region of space it occupies in a fixed reference configuration, and denote by $\bfx_\mat$ an arbitrary material point of $\calB_\mat$. The referential body $\calB_\mat$ then undergoes a motion $\bfx = \bfchi(\bfx_\mat,t)$ to the deformed body $\calB_t$ with deformation gradient given by\footnote{Following common notation \citep{gurtin2010mechanics}, the symbols $\nabla$ and $\Div$ denote the gradient and divergence with respect to the material point $\bfx_\mat$ in the reference configuration; while $\grad$ and $\divx$ denote these operators with respect to the point $\bfx = \bfchi(\bfx_\mat,t)$ in the deformed configuration.
Also, we write $\Tr \bfA$, $\sym \bfA$, $\skw \bfA$, and $\bfA_{0}$ respectively, for the trace, symmetric, skew, and deviatoric parts of a tensor $\bfA$. Lastly, the inner product of tensors $\bfA$ and $\bfB$ is denoted by $\bfA \tendot \bfB$, and the magnitude $\bfA$ by $\left| \bfA \right| = \sqrt{\bfA \tendot \bfA}$.} 
\begin{equation}\label{eqn:DefGrad}
  \bfF = \nabla \bfchi, \quad\text{such that}\quad J=\det\bfF>0. 
\end{equation}
The right and left Cauchy-Green deformation tensors are given by
\begin{equation}
\bfC=\bfF^{\trans}\bfF \,,
\end{equation} 
and
\begin{equation}
\bfB=\bfF\bfF^{\trans} \,.
\end{equation}
Also, the polar decomposition of the deformation gradient
\begin{equation}
  \bfF = \bfR\bfU
\end{equation}
allows its split into a rotation $\bfR$, and a symmetric stretch $\bfU$.

\subsection{Free energy}
Following recent literature, we introduce a scalar damage field $d \in [0,1]$, which characterizes an intact (damage-free) state when $d = 0$ and a fully-damaged state when $d = 1$. Consequently, we introduce the degraded free energy density per unit reference volume due to damage evolution such that 
\begin{equation}
\begin{split}
\hat{\psi}_\mat & = \hat{\psi}_\mat(\bfC,d, \Grad d) \\
 & = \hat{\psi}_\mat^{\ast}(\bfC,d) + \hat{\psi}_{\mat,\text{nonlocal}}(\Grad d) \,,
\label{eqn:FreeEnergyGeneric}
\end{split}
\end{equation}
where $\hat{\psi}_\mat^{\ast}$, $\hat{\psi}_{\mat,\text{nonlocal}}$, and $\Grad d$ represent the degraded free energy, the nonlocal damage contribution of the free energy, and the \emph{referential} gradient of damage defined by $\Grad d \Def \frac{\partial d}{\partial \bfx_\mat}$,
respectively. 
The degraded free energy is defined such that
\begin{equation}
 \hat{\psi}_\mat^{\ast} \Def g(d) \hat{\psi}_\mat^{0} (\bfC)\,,
\label{eqn:LocalFreeEnergyGeneric}
\end{equation}
where the monotonically decreasing degradation function $g(d)$ has the following properties

\begin{equation}
    g'(d) \leq 0 \quad \text{with} \quad g(0) = 1, \quad g(1) = 0, \quad \text{and} \quad g'(1) = 0\,,
\end{equation}
and is constitutively prescribed as
\begin{equation}\label{eqn:g(d)}
    g(d) = (1 - d)^2 \,,
\end{equation} and $\hat{\psi}_\mat^{0} (\bfC)$ is the free energy of the undamaged body. Next, the nonlocal damage contribution of the free energy is defined such that 

\begin{equation}
    \hat{\psi}_{\mat,\text{nonlocal}} (\Grad d) \Def \frac{1}{2} \epsilonR l^2 |\Grad d|^2\,,
       \label{eqn:NonLocalFreeEnergy}
\end{equation}
where $\epsilonR$ and $l$ correspond to the fracture energy per unit reference volume and the intrinsic characteristic length scale for the damage process, respectively.

\subsection{Governing equations and boundary conditions in the referential configuration}

\subsubsection{Referential macroforce balance: Balance of forces and moments}

As is standard, neglecting inertial effects, the balance of forces and moments in the reference body $\Br$  are  expressed as 
\begin{equation}\label{eqn:BalanceForcesMoments_Referential}
  \divx\bfT_\mat  + \bfb_\mat = \zed\quad\text{and}\quad\bfT_\mat \bfF^\trans = \bfF \bfT_\mat^{\trans},
\end{equation}
where $\bfT_\mat$ and $\bfb_\mat$ represent the Piola-Kirchoff stress and an external body force per unit reference volume, respectively.
Additionally, the standard boundary conditions are prescribed motion and tractions
\begin{equation}\label{eqn:BalanceForcesMoments_Referential_BC}   
  \left.
  \begin{aligned}
 \bfchi = \breve{\bfchi} \qquad  \qquad &\text{on} \quad \Schi\,,\\
       \bfT_\mat \bfn_\mat = \bf{\breve{t}_\mat} \qquad  \qquad &\text{on} \quad \Stmat\,,
  \end{aligned}  
  \right\}
\end{equation}
where $\Schi$ and $\Stmat$ represent complementary subsurfaces of the boundary $\partial\Br$ of the body $\Br$, such that $\partial\Br = \Schi \cup \Stmat$ and $\Schi \cap \Stmat = \varnothing$, with an initial condition $\bfchi (\bf\bfx_\mat,0) = \bfchi_{0} (\bf\bfx_\mat)$ in $\Br$.

\subsubsection{Referential microforce balance: Damage evolution equation}
The balance of microforces in the reference body $\Br$  takes the form 
\begin{equation}\label{eqn:BalanceMicroForces_Referential}
  \Div \xiR  - \omegaR = 0\,,
\end{equation}
where the scalar and vector microscopic stresses, $\omegaR$ and $\bf{\xiR}$, conjugates to $d$ and $\Grad d$, respectively, are constitutively prescribed such that
\begin{equation}
\label{eqn:Constitutive_MicroStresses_Referential}
\begin{split}
    \omegaR & = \frac{\partial \hat{\psi}_\mat^{\ast}}{\partial d} + \omega_{\mat, \text{diss}}\,, \quad \text{and} \\
\bf{\xiR} & = \frac{\partial  \hat{\psi}_{\mat,\text{nonlocal}}}{\partial \Grad d} = \epsilonR l^2 \Grad d \,.
\end{split}
\end{equation}
Moreover, the dissipative part of $\omegaR$ is given by
\begin{equation}
\label{eqn:Constitutive_MicroStressVisco_Referential}
\omega_{\mat, \text{diss}} = \alpha  + \zetaR \ddot\,,
\end{equation}
with $\alpha$ representing the dissipated fracture energy per unit reference volume during damage evolution such that $\alpha = \epsilonR d$, and $\zetaR$ representing a viscous regularization parameter.
Next, use of \eqref{eqn:g(d)}, \eqref{eqn:Constitutive_MicroStresses_Referential} and \eqref{eqn:Constitutive_MicroStressVisco_Referential} allow \eqref{eqn:BalanceMicroForces_Referential} to be recast such that
\begin{equation}\label{eqn:DamageEvolution1}
    \zetaR \ddot = 2 (1-d) \hat{\psi}_\mat^{0} + \epsilonR l^2 \Div(\Grad d) - \epsilonR d \,,
\end{equation}
along with boundary conditions such that
\begin{equation}\label{eqn:BalanceMicroForces_Referential_BC}   
  \left.
  \begin{aligned}
 d = 0 \qquad  \qquad &\text{on} \quad \Sd\,,\\
       \Grad d \cdot \bfn_\mat = 0  \qquad &\text{on} \quad \SxiR\,,
  \end{aligned}  
  \right\}
\end{equation}
where $\Sd$ and $\SxiR$ represent complementary subsurfaces of the boundary $\partial\Br$ of the body $\Br$, such that $\partial\Br = \Sd \cup \SxiR$ and $\Sd \cap \SxiR = \varnothing$,  with an initial condition $d (\bfx_\mat,0) = 0$ in $\Br$.
Lastly, to enforce damage irreversibility ($\ddot > 0 $), we introduce the monotonically increasing history function following \citet{miehe2010phase} such that 
\begin{equation}\label{eqn:HistoryFunction}
    \HR (t) = \max_{s \in [0,t]}  \hat{\psi}_{\mat}^{0} (\bfC(s))\,,
\end{equation}
which allows recasting \eqref{eqn:DamageEvolution1} into
\begin{equation}\label{eqn:DamageEvolution_Referential}
        \zetaR \ddot = 2 (1-d) \HR - \epsilonR d + \epsilonR l^2 \Div (\Grad d) \,.
\end{equation}

\subsection{Governing equations and boundary conditions in the spatial configuration}

\subsubsection{Spatial macroforce balance: Balance of forces and moments}

Once again, neglecting inertial effects, the balance of forces and moments in the deformed body $\Bt$ are expressed as 
\begin{equation}\label{eqn:BalanceForcesMoments_Spatial}
  \divx\bfT  + \bfb = \zed\quad\text{and}\quad\bfT = \bfT^{\trans},
\end{equation}
where $\bfT$ and $\bfb$ represent the Cauchy stress and an external body force per unit deformed volume, respectively. Moreover, the standard boundary conditions are prescribed displacement and tractions
\begin{equation}\label{eqn:BalanceForcesMoments_Spatial_BC}   
  \left.
  \begin{aligned}
       \bfu = \breve{\bfu} \qquad  \qquad &\text{on} \quad \Su\,,\\
       \bfT \bfn = \bf{\breve{t}} \qquad  \qquad &\text{on} \quad \St\,,
  \end{aligned}  
  \right\}
\end{equation}
where $\Su$ and $\St$ represent complementary subsurfaces of the boundary $\partial\Bt$ of the body $\Bt$, such that, $\partial\Bt = \Su \cup \St$ and $\Su \cap \St = \varnothing$, with an initial condition $\bfu (\bfx_\mat,0) = \bfu_{0} (\bfx_\mat)$ in $\Br$.

\subsubsection{Spatial microforce balance: Damage evolution equation}

The balance of microforces in the deformed body $\Bt$  takes the form 
\begin{equation}\label{eqn:BalanceMicroForces_Spatial}
  \divx \bfxi  - \omega = 0\,,
\end{equation}
 where the scalar and vector microscopic 
 stresses, $\omega$ and $\bfxi$, conjugates to $d$ and $\grad d$, respectively, are related to their referential counterparts following
\begin{equation}
\label{eqn:Constitutive_MicroStresses_Spatial}
\begin{split}
    \omega & = J^{-1} \omegaR\,, \quad \text{and} \\
{\bfxi} & = J^{-1} \bfF \bf{\xiR}\,,
\end{split}
\end{equation}
which allows \eqref{eqn:Constitutive_MicroStresses_Referential} to be recast such that 
\begin{equation}
\label{eqn:Constitutive_MicroStresses_Spatial_Recast}
\begin{split}
    \omega & = J^{-1} \frac{\partial \hat{\psi}_\mat^{\ast}}{\partial d} + \epsilonS d + \zeta \ddot \,, \quad \text{and} \\
{\bfxi} &  = J^{-1} \epsilonR l^2 \bfF \bfF^\trans \grad d \,,
\end{split}
\end{equation}
with $\epsilonS = J^{-1} \epsilonR$ and $\zeta = J^{-1} \zetaR$. 

Next, use of \eqref{eqn:g(d)}, \eqref{eqn:HistoryFunction} and \eqref{eqn:Constitutive_MicroStresses_Spatial_Recast}
allows \eqref{eqn:BalanceMicroForces_Spatial} to be recast such that
\begin{equation}\label{eqn:DamageEvolution_Spatial}
\zeta \ddot =   2 (1 - d) \HS -  \epsilonS d + \epsilonR l^{2} \divx (J^{-1} \bfB \grad d)\,,
\end{equation}
where the spatial history function is related to its referential counterpart using $\HS = J^{-1} \HR$.
Lastly, the boundary conditions are given by 
\begin{equation}\label{eqn:BalanceMicroForces_Spatial_BC}   
  \left.
  \begin{aligned}
 d = 0 \qquad  \qquad &\text{on} \quad \Sd\,,\\
       \grad d \cdot \bfn = 0  \qquad &\text{on} \quad \Sxi\,,
  \end{aligned}  
  \right\}
\end{equation}
where $\Sd$ and $\Sxi$ represent complementary subsurfaces of the boundary $\partial\Bt$ of the body $\Bt$, such that, $\partial\Bt = \Sd \cup \Sxi$ and $\Sd \cap \Sxi = \varnothing$, with an initial condition $d (\bfx_\mat,0) = 0$ in $\Br$.

\section{Finite element implementation}\label{sec:FE_Implementation}

In this section, we explore the similarities between the heat equation and phase-field evolution by examining an Abaqus finite element implementation. We then comment on its validity and provide an Abaqus implementation of the gradient-damage theory for fracture in materials undergoing large deformation using user material subroutines UMAT and UMATHT.

\subsection{Abaqus implementation: Analogy to the heat equation}

As per the Abaqus documentation \citep{AbqStan}, the heat equation (energy balance) is given by

\begin{equation}\label{eqn:HeatEquation_Abaqus}
\int_V \rho \dot{U}\,dV = \int_S q\,dS + \int_V r\,dV\,,
\end{equation}
where $V$ is the volume of solid material with surface area $S$, $\rho$ is the density of the material, $\dot{U}$ is the material time rate of the internal thermal energy, $q$ is the heat flux per unit area flowing into the body, and $r$ is the heat supplied externally into the body per unit volume.
Using the divergence theorem, along with Fourier's law and the important relation $\dot{U} \Def C \dot{\theta}$, with $C$ representing the specific heat measured in energy per unit mass per temperature for a fixed deformation, and $\dot{\theta}$ the rate of change of temperature, the heat equation \eqref{eqn:HeatEquation_Abaqus} may be recast into its strong form 
\begin{equation} \label{eq:HeatEquation_StrongForm}
    \rho C \dot{\theta} = \divx (\bfK \grad \theta) + r\,,
\end{equation}
where $\bfK$ is the thermal conductivity tensor.

It is important to note here that Abaqus formulates the heat equation in the spatial (deformed) configuration in the sense that it uses spatial operators ``$\divx$'' \underline{rather than} ``$\Div$'' and ``$\grad$'' \underline{rather than} ``$\Grad$''. This makes the damage evolution equation in the referential configuration \eqref{eqn:DamageEvolution_Referential} \emph{fundamentally different} from the heat equation used in the Abaqus UMAT subroutine when dealing with large deformation kinematics.
When applied within the \emph{limits of small deformation} kinematics (i.e., brittle failure in linear elastic solids), since both referential and spatial operators are essentially the same, using the heat equation as provided in UMAT is an appropriate approximation.

Consequently, for an accurate implementation of the gradient-damage theory for fracture in materials undergoing large deformation using Abaqus user material subroutines, one needs to apply the analogy between the heat equation and the damage evolution equation in its spatial configuration, i.e., \eqref{eqn:DamageEvolution_Spatial}.

We start by recasting \eqref{eqn:DamageEvolution_Spatial} such that
\begin{equation}\label{eqn:DamageEvolution_Spatial_RECAST}
    J^{-1} \zetaR \ddot  =  \epsilonR l^{2} \divx (J^{-1} \bfB \grad d) + \underbrace{2 (1 - d) J^{-1} \HR -  J^{-1} \epsilonR d}_r\,,
\end{equation}
and use it in Table \ref{tab:comparison} to show its resemblance to the heat equation using a term-by-term comparison.
\begin{table}[h]
    \centering
    \renewcommand{\arraystretch}{1.5} 
    \begin{tabular}{c|c|c|c}
       \textbf{Equation}  & \textbf{Transient term} & \textbf{Conduction term} & \textbf{Source term} \\ \hline
        \textbf{ \eqref{eq:HeatEquation_StrongForm}} & $\rho C \dot{\theta}$ & $\divx (\bfK \grad \theta)$ & $r$ \\ 
        \textbf{ \eqref{eqn:DamageEvolution_Spatial_RECAST}} & $J^{-1} \zetaR \ddot$ & $ \epsilonR l^{2} \divx (J^{-1} \bfB \grad d)$ & $2 (1 - d) J^{-1} \HR -  J^{-1} \epsilonR d$ \\
    \end{tabular}
    \caption{Term-by-term comparison of the heat equation \eqref{eq:HeatEquation_StrongForm} and the damage evolution equation \eqref{eqn:DamageEvolution_Spatial_RECAST}.}
    \label{tab:comparison}
\end{table}

In order to use the heat equation provided by Abaqus through the user subroutine UMAT, one needs to modify each of the terms in Table \ref{tab:comparison} as follows, with the caveat that the damage ``$d$'' is represented by temperature ``$\theta$'':

\begin{itemize}
    \item Starting with the transient term, one needs to impose the equality $\rho C  = J^{-1} \zetaR$, which requires access to the thermal constitutive behavior that can be achieved using the user subroutine ``UMATHT.'' The details are provided in Appendix \ref{Appendix:Transient}.
    \item Similarly, one needs to make two changes to the conduction term such that:
\begin{enumerate}
    \item The thermal conductivity tensor $\bfK$ in Fourier's law must be replaced by $J^{-1} \bfB$.
    \item The conduction term must be multiplied by a pre-factor $\epsilonR l^{2}$.
\end{enumerate}
Both changes can be also done through the user subroutine ``UMATHT.'' The details are provided in Appendix \ref{Appendix:Conduction}.
\item Lastly, the source term needs to be $r = 2 (1 - d) J^{-1} \HR -  J^{-1} \epsilonR d$, which can be directly achieved through the ``RPL'' functionality in the user subroutine ``UMAT.'' The details are provided in Appendix \ref{Appendix:Source}.
\end{itemize}

\subsection{Verification of implementation}
In this section, we verify our proposed implementation in \cite{AbqStan} by examining a series of simple boundary value problems, chosen to decouple each term as much as possible for detailed verification.
We first study the homogeneous deformation of a rubber-like material subjected to large-stretch uniaxial tension at different stretch rates. This example serves to verify the coupled transient damage initiation/evolution only due to a source term, without the presence of spatial gradients. 
We then study the inhomogeneous damage propagation in a rubber-like material in which we apply damage to a given geometry and allow it to propagate.\footnote{While this scenario may not reflect a typical boundary value problem, it provides a valuable and controlled setting to rigorously verify our implementation by isolating and analyzing individual contributions.} This example serves to verify damage propagation throughout the material.
In both examples, we compare our implementation using UMAT+UMATHT against the UEL by \citet{lee2023finite}, where we adopt the slightly compressible Neo-Hookean material model with the constitutive details provided in Appendix \ref{Appendix:Constitutive_Neo} for completeness.

\subsubsection{Homogeneous damage initiation/evolution in a rubber-like material}\label{sec:HomogeneousDamage}

\begin{figure}[htb]
\centering
	 \includegraphics[width = .35\textwidth]{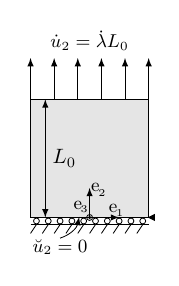}
\caption{Schematic of the boundary value problem of a rubber-like material subjected to large-stretch uniaxial tension at different stretch rates.}
\label{fig:Schematic_BVP1}
\end{figure}

\begin{figure}[htb]
\centering
\begin{tabular}{cc}
\includegraphics[width = .5\textwidth]{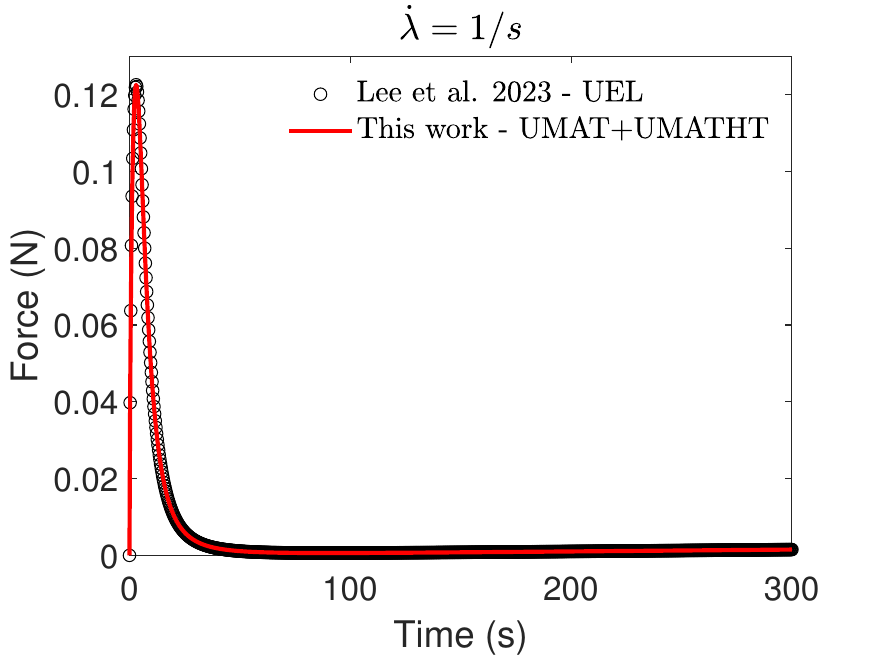} &
\includegraphics[width = .5\textwidth]{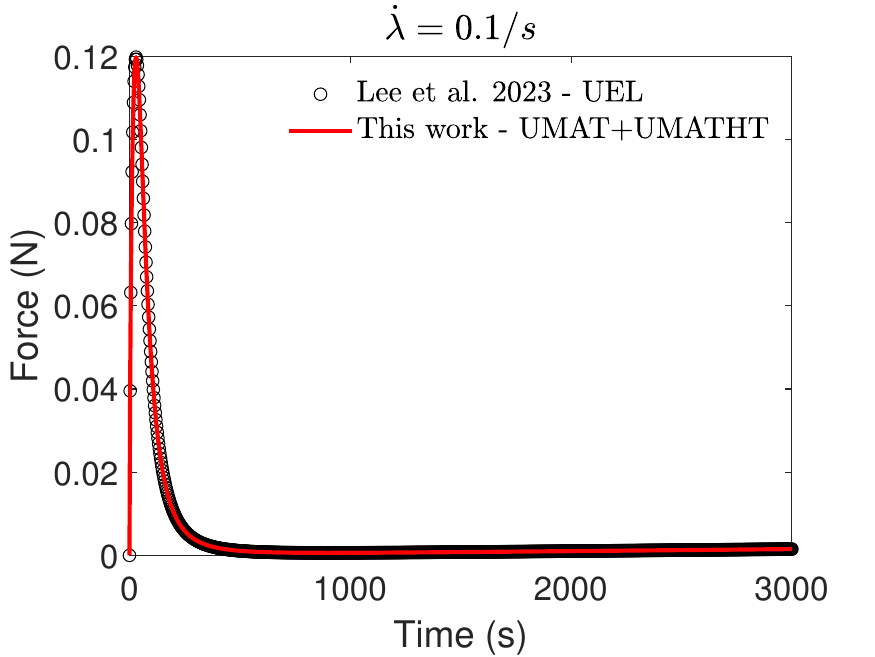}\\
(a) & (b)
\end{tabular}
\begin{tabular}{cc}
\includegraphics[width = .5\textwidth]{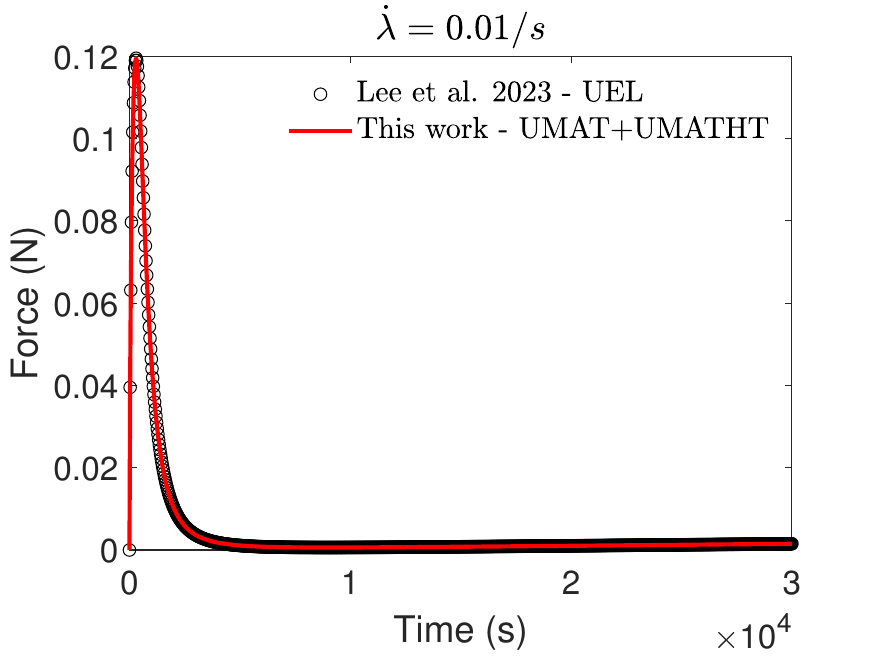}\\
(c)
\end{tabular}
    \caption{Comparison of the force-time response between the user element (UEL) by \citet{lee2023finite} and Abaqus user material subroutines (UMAT+UMATHT) developed in this work, at multiple stretch rates.}
\label{fig:BVP1_FT}
\end{figure}

\begin{figure}[htb]
\centering
\begin{tabular}{cc}
\includegraphics[width = .5\textwidth]{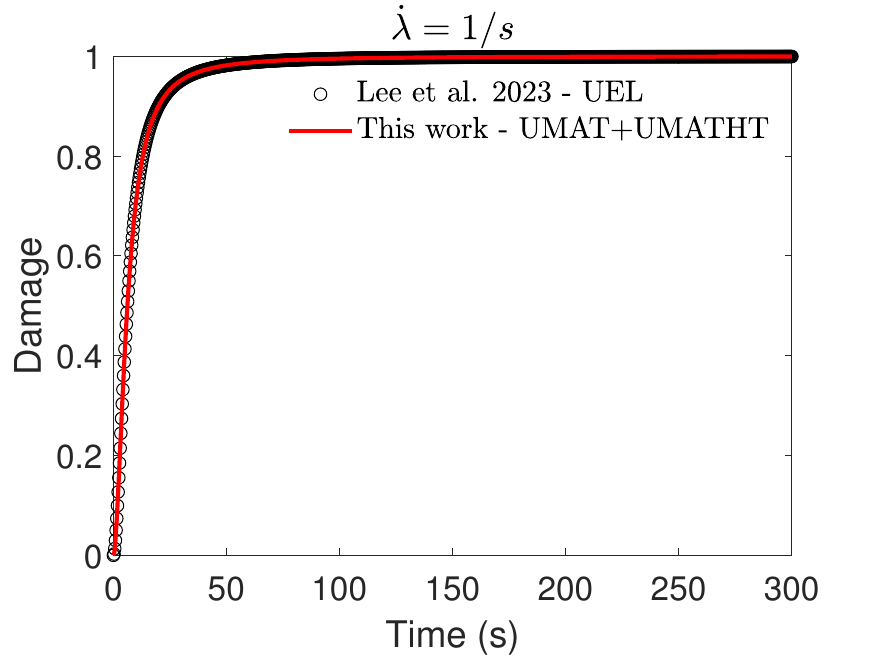} &
\includegraphics[width = .5\textwidth]{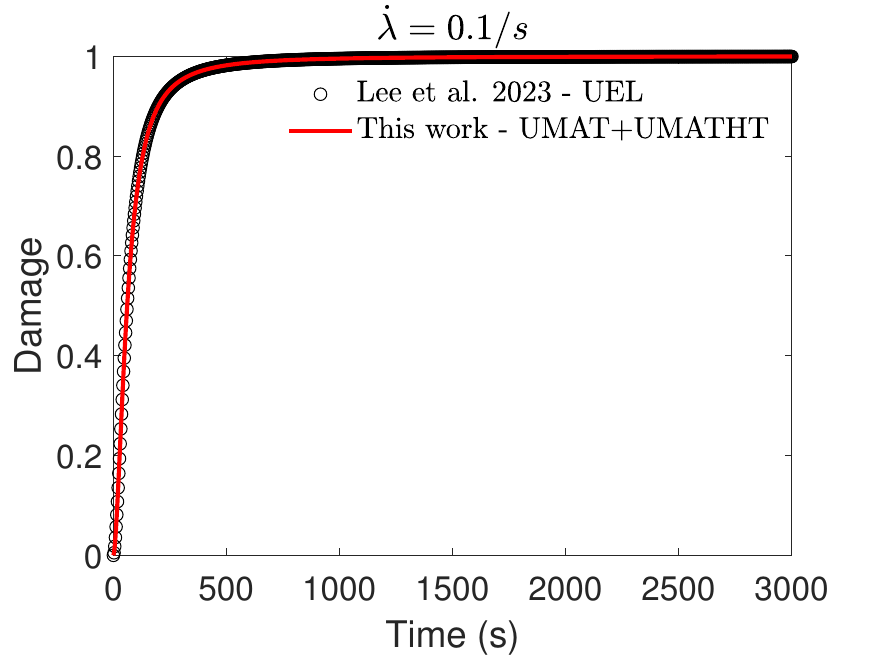}\\
(a) & (b)
\end{tabular}
\includegraphics[width = .5\textwidth]{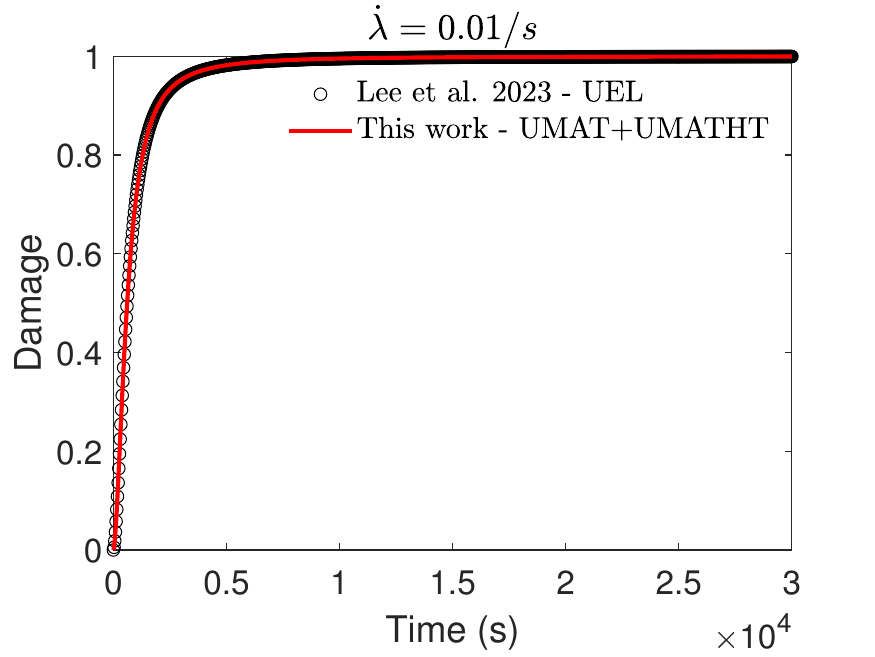}\\
(c) 
\caption{Comparison of the damage-time response between the user element (UEL) by \citet{lee2023finite} and Abaqus user material subroutines (UMAT+UMATHT) developed in this work, at multiple stretch rates.}
\label{fig:BVP1_DT}
\label{fig:BVP1}
\end{figure}

In this example, we consider the plane strain homogeneous deformation in a rubber-like material subjected to a uniaxial tension at multiple stretch rates, as shown in Figure \ref{fig:Schematic_BVP1}.
Specifically, the reference body $\Br$ is a solid rectangle of initial dimension $L_0 = 0.01\,$mm, with a roller boundary condition applied to its bottom surface, and its top surface subjected to a velocity ($\dot{u}$) boundary condition for various stretch rates ($\dot{\lambda}$), which can be achieved using the standard kinematic relation, $\dot{u} \Def \dot{\lambda} L_0$. Both the left and right sides are traction-free. Consequently, we modeled this problem using a single 4-node linear displacement-temperature plane strain element (CPE4T).
For this boundary value problem, due to the spatial homogeneity, the only active component in the damage evolution equation, \eqref{eqn:DamageEvolution_Spatial_RECAST}, is the source term ``$r$'' such that
\begin{equation}\label{eqn:DamageEvolution_Spatial_BVP1}
    J^{-1} \zetaR \ddot  = \underbrace{2 (1 - d) J^{-1} \HR -  J^{-1} \epsilonR d}_r\,.
\end{equation}
Figures \ref{fig:BVP1_FT} and \ref{fig:BVP1_DT} compare the results of our implementation using Abaqus user material subroutines (UMAT+UMATHT), and the user elements (UEL) by \citet{lee2023finite}.

Specifically, Figure \ref{fig:BVP1_FT} shows the force-time response for multiple stretch rates, while Figure \ref{fig:BVP1_DT} shows the damage-time response, for multiple stretch rates. It can be observed from both Figures that our implementation here shows an identical response when compared to the results of the UEL by \citet{lee2023finite}, confirming the validity of the implementation in capturing the \emph{transient} damage initiation and evolution due to a source term ``$r$'' as described in \eqref{eqn:DamageEvolution_Spatial_BVP1}. We also observe slight variations in the response at different loading rates, which can be attributed to the rate-dependent nature of \eqref{eqn:DamageEvolution_Spatial_BVP1}.

\subsubsection{Inhomogeneous damage propagation in a rubber-like material}\label{sec:NonHomogeneousDamage}
In this example, we consider inhomogeneous damage propagation in a plane strain rubber-like material subjected to various boundary conditions. Specifically, we consider two scenarios:
\begin{itemize}
    \item Scenario 1: We consider a fixed rectangular geometry with prescribed damage along the left and top sides. Specifically, the damage, $d$, is ramped linearly from 0 to 0.95 over 100 seconds, then remains fixed for another 200 seconds, allowing its propagation over a total of 300 seconds, as seen in Figure \ref{fig:Schematic_BVP2_Scenario1}.
\begin{figure}[htb]
\centering
	 \includegraphics[width = .4\textwidth]{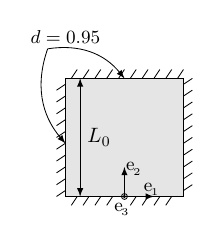}
\caption{Schematic of the boundary value problem studied in Scenario 1, showing the inhomogeneous damage propagation in a fixed rubber-like material.}
\label{fig:Schematic_BVP2_Scenario1}
\end{figure}
    \item Scenario 2: Two steps are considered in this scenario. In the first step, damage, $d$, is ramped linearly from 0 to 0.45 over 100 seconds along the left and top surfaces of the rectangle while it is fixed. In the second step, a displacement boundary condition is applied to its right ($u_1$) and bottom surfaces ($u_2$) simultaneously, ramped linearly from 0 to 0.05 mm over 200 seconds, while the applied damage, $d$, applied in the first step is held constant throughout the entire step, as seen in Figure \ref{fig:Schematic_BVP2_Scenario2}.
\begin{figure}[htb]
\centering
	 \includegraphics[width = .65\textwidth]{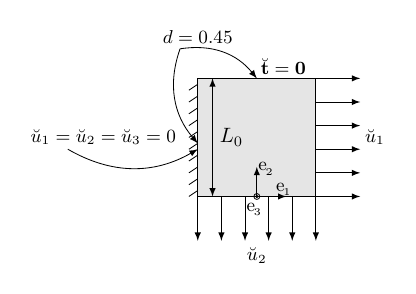}
\caption{Schematic of the boundary value problem studied in Scenario 2, showing the inhomogeneous damage propagation in a deforming rubber-like material.}
\label{fig:Schematic_BVP2_Scenario2}
\end{figure}
\end{itemize}
We note that the same geometry is used in the two scenarios, consisting of a reference body $\Br$ that is a solid rectangle of initial dimension $L_0 = 0.1\,$mm, which was discretized using 10000 4-node linear displacement-temperature plane strain elements (CPE4T) with element size $l_e = 0.001\,$mm.

\begin{figure}[htb]
\centering
\begin{tabular}{cc}
\qquad \qquad \citet{lee2023finite} - UEL  & \qquad \qquad This work - UMAT+UMATHT\\
\includegraphics[width = .45\textwidth]{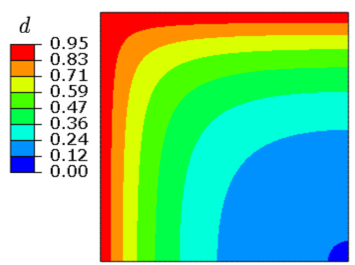} &
\includegraphics[width = .45\textwidth]{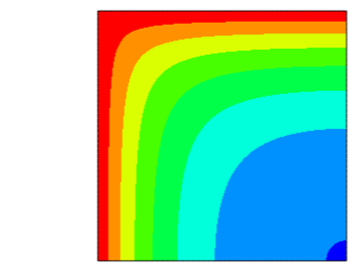}\\
\qquad \qquad (a) & \qquad \qquad (b)
\end{tabular}
\caption{Comparison of the damage propagation in a fixed rectangular geometry between the user element (UEL) by \citet{lee2023finite} and Abaqus user material subroutines (UMAT+UMATHT) developed in this work.}
\label{fig:BVP2_Scenario1}
\end{figure}

\begin{figure}[htb]
\centering
\begin{tabular}{cc}
\qquad \qquad \citet{lee2023finite} - UEL  & \qquad \qquad This work - UMAT+UMATHT\\
\includegraphics[width = .45\textwidth]{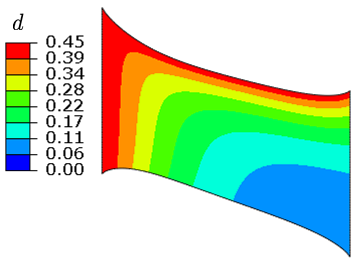} &
\includegraphics[width = .45\textwidth]{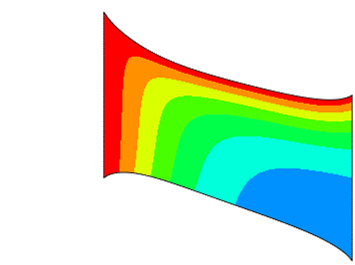}\\
\qquad \qquad (a) & \qquad \qquad (b)
\end{tabular}
\caption{Comparison of the damage propagation in a deforming rectangular geometry between the user element (UEL) by \citet{lee2023finite} and Abaqus user material subroutines (UMAT+UMATHT) developed in this work.}
\label{fig:BVP2_Scenario2}
\end{figure}

\begin{Remark}
Once again, we note that although these scenarios do not reflect typical boundary value problems, their simplicity provides a valuable and controlled setting to rigorously verify our implementation by isolating and analyzing individual contributions.
\end{Remark}
\paragraph{Scenario 1: Damage propagation in a fixed rectangular geometry\\}
In this first scenario, since the body is fixed (i.e., $\bfF = \id$, $\bfB = \id$, and $J = 1$), the damage evolution equation \eqref{eqn:DamageEvolution_Spatial_RECAST} can be decoupled such that only the ``conduction'' term is active
\begin{equation}\label{eqn:PP_PDE_Scenario1}
     \zetaR \ddot  =  \epsilonR l^{2} \divx (  \grad d)\,.    
\end{equation}
Figure \ref{fig:BVP2_Scenario1} shows the simulation results after 300 seconds, where we can see an excellent agreement between our implementation using UMAT+UMATHT and the UEL by \citet{lee2023finite}.  This confirms the validity of our UMAT+UMATHT implementation in capturing the damage propagation due to ``conduction'' as described in \eqref{eqn:PP_PDE_Scenario1}. It is worth mentioning that although not shown here, this agreement was maintained throughout the entire simulation.
\paragraph{Scenario 2: Damage propagation in a deforming rectangular geometry \\}
In the second scenario, the deforming body allows for testing the fully coupled damage evolution equation \eqref{eqn:DamageEvolution_Spatial_RECAST}, verifying the robustness of our implementation. 
Similarly, Figure \ref{fig:BVP2_Scenario2} shows the simulation results after 300 seconds, where we can see an excellent agreement between our implementation using UMAT+UMATHT and the UEL by \citet{lee2023finite}. This confirms the validity of our implementation in capturing the fully coupled damage propagation in materials undergoing large deformation. Once again, this agreement was maintained throughout the entire simulation.

To summarize, following the results from Sections \ref{sec:HomogeneousDamage} and \ref{sec:NonHomogeneousDamage}, we have verified our finite element implementation of a large deformation gradient-damage theory for fracture using Abaqus user material subroutines UMAT and UMATHT, by comparing our implementation against the UEL by \citet{lee2023finite}. We use it in what follows to solve classical boundary value problems often encountered in the phase-field literature.
\clearpage

\section{Applications to fracture problems}\label{sec:Applications}
In this section, we show the capabilities of our gradient-damage implementation for large deformation fracture using Abaqus user material subroutines UMAT and UMATHT. We validate it against key benchmark problems from the literature \citep{miehe2010phase, miehe2014phase}. Specifically, we consider the following examples:
\begin{itemize}
    \item Penny-shaped 3D specimen in tension utilizing a large deformation rubber material constitutive model.
    \item Single edge notched tension (SENT) test utilizing a small deformation linear elastic material constitutive model.
    \item Single edge notched beam (SENB) test utilizing a large deformation rate-dependent plasticity material constitutive model.
\end{itemize}

\subsection{Penny-shaped specimen in tension: large deformation rubber}
\begin{figure}[htb]
\centering
	 \includegraphics[width = 0.7\textwidth]{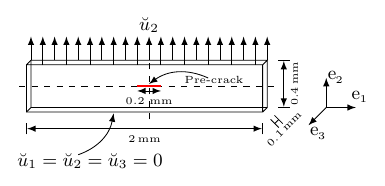}
\caption{Schematic showing the geometry and boundary conditions for the penny-shaped specimen in tension.}
\label{fig:Schematic_PennyShaped}
\end{figure}

In this example, we examine the fracture response of a penny-shaped specimen with dimensions $L_0 = 2\,$mm, $W_0 = 0.4\,$mm, and $t_0 = 0.1\,$mm. The specimen contains a pre-crack of length $a = 0.2\,$mm at its center, as shown in Figure \ref{fig:Schematic_PennyShaped}.
The bottom surface is fixed, and displacement is applied to its top surface, such that $u_2$ increases linearly to $0.64\,$mm over $1600$ seconds. All other surfaces are traction-free. The material constitutive response is taken as a compressible Neo-Hookean solid, with constitutive details and material parameters provided in Appendix \ref{Appendix:Constitutive_Neo}.
Due to the symmetry, we only simulate one-eighth of the geometry. Specifically, the mesh consists of 6906 8-node linear displacement-temperature elements (C3D8T) with 2 elements through the (half-)thickness, with the elements near the pre-crack path having a mesh size $l_e = 0.005\,$mm.
Additionally, we use the UEL provided by \citet{lee2023finite} to simulate the same boundary value problem for direct comparison.

\begin{figure}[htb]
\centering
	 \includegraphics[width = 0.5\textwidth]{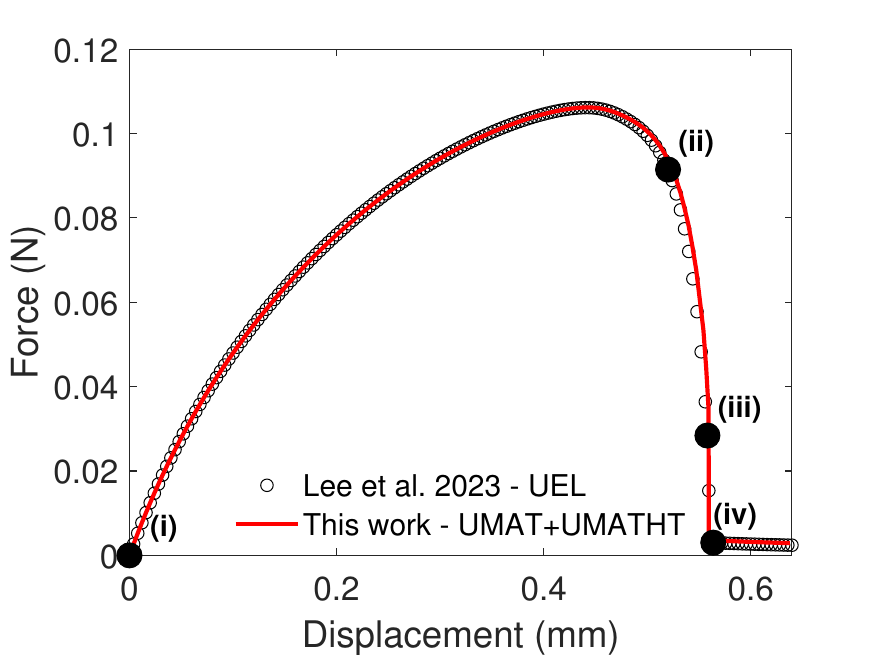}
\caption{Comparison of the force-displacement results for the penny-shaped specimen in tension between the user element (UEL) by \citet{lee2023finite} and Abaqus user material subroutines (UMAT+UMATHT) developed in this work.}
\label{fig:Penny_F_D}
\end{figure}

\begin{figure}[htb]
\centering
\begin{tabular}{c}
\includegraphics[width = .65\textwidth]{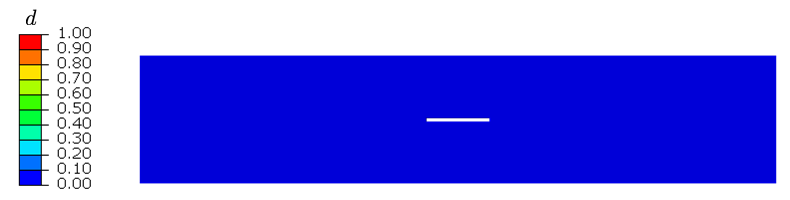} \\
(i) \\
\end{tabular}
\begin{tabular}{cc}
\qquad \citet{lee2023finite} - UEL  & \qquad  This work - UMAT+UMATHT\\
\includegraphics[width = .33\textwidth]{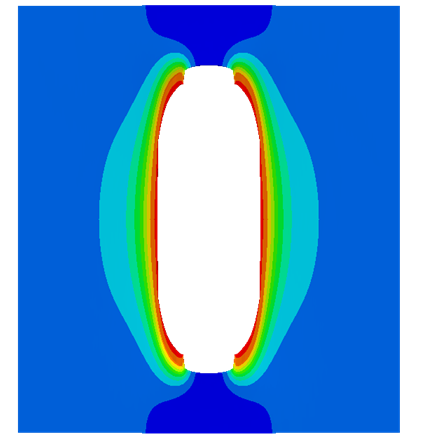} &
\includegraphics[width = .33\textwidth]{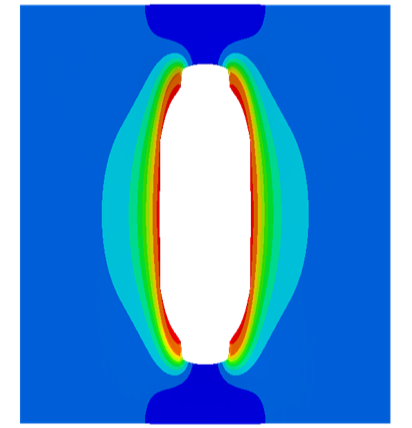}\\
 (ii) &  (ii) \\
\includegraphics[width = .33\textwidth]{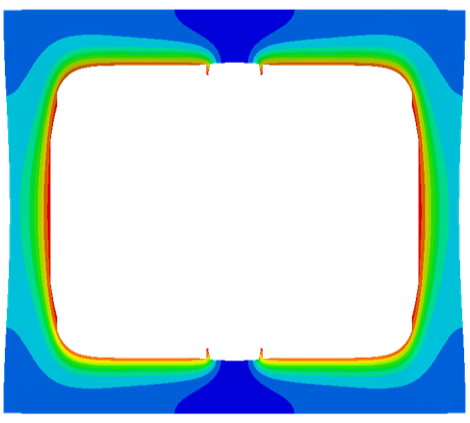} &
\includegraphics[width = .33\textwidth]{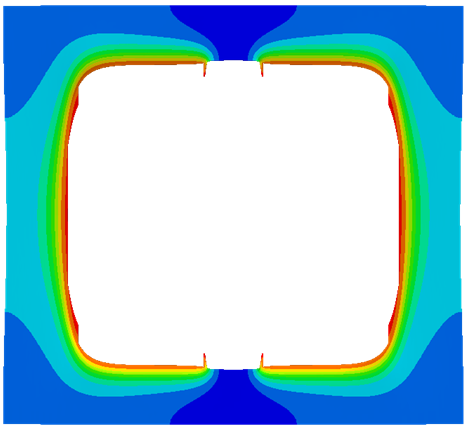}\\
 (iii) &  (iii) \\
\includegraphics[width = .33\textwidth]{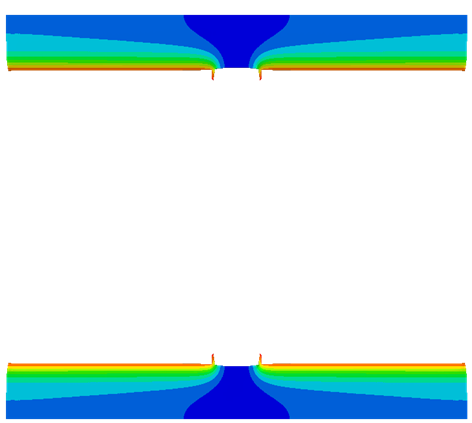} &
\includegraphics[width = .33\textwidth]{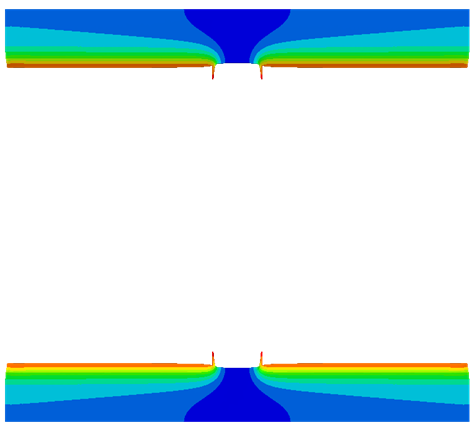}\\
\qquad (iv) & \qquad (iv) \\
\end{tabular}
\caption{Comparison of the damage propagation for the penny-shaped specimen in tension between the user element (UEL) by \citet{lee2023finite} and Abaqus user material subroutines (UMAT+UMATHT) developed in this work at different instants shown in Figure \ref{fig:Penny_F_D}. Note that the elements are not shown for $d > 0.95$.}
\label{fig:Penny_Damage_Evolution}
\end{figure}

It can be observed from Figure \ref{fig:Penny_F_D} that the force-displacement response is in excellent agreement between our implementation and the UEL by \citet{lee2023finite}.
Moreover, Figure \ref{fig:Penny_Damage_Evolution} shows the damage evolution over time. The fracture initiates near the pre-crack and is propagated through the specimen until complete rupture, similar to what was reported in \citet{lee2023finite}. This benchmark problem proves the validity of the implementation developed in this work using UMAT+UMATHT in predicting the fracture response in materials undergoing large deformation.
\clearpage
\subsection{Single edge notched tension test: small deformation linear elasticity}
\begin{figure}[htb]
\centering
	 \includegraphics[width = 0.45\textwidth]{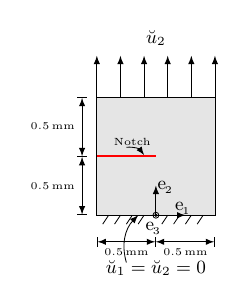}
\caption{Schematic showing the geometry and boundary conditions for the single edge notched tension specimen.}
\label{fig:Schematic_SENT}
\end{figure}

In this second example, we examine the fracture response of a single edge notched square plate of edge length $L_0 = 1\,$mm, with a horizontal notch at mid-height, extending from the left outer surface to the center of the specimen, as shown in Figure \ref{fig:Schematic_SENT}.
The bottom surface is fixed, and displacement is applied to its top surface, such that $u_2$ increases linearly to $0.0035\,$mm over $100$ seconds, resulting in a Mode I fracture. All other surfaces are traction-free. The material constitutive response is taken as a linear elastic solid, with details and material parameters provided in Appendix \ref{Appendix:Constitutive_LE}.
We model this problem using the plane strain approximation. Specifically, the mesh consists of 5289 4-node linear displacement-temperature elements (CPE4T), with the elements along the notch path having a mesh size $l_e = 0.005\,$mm.

\begin{figure}[htb]
\centering
	 \includegraphics[width = 0.5\textwidth]{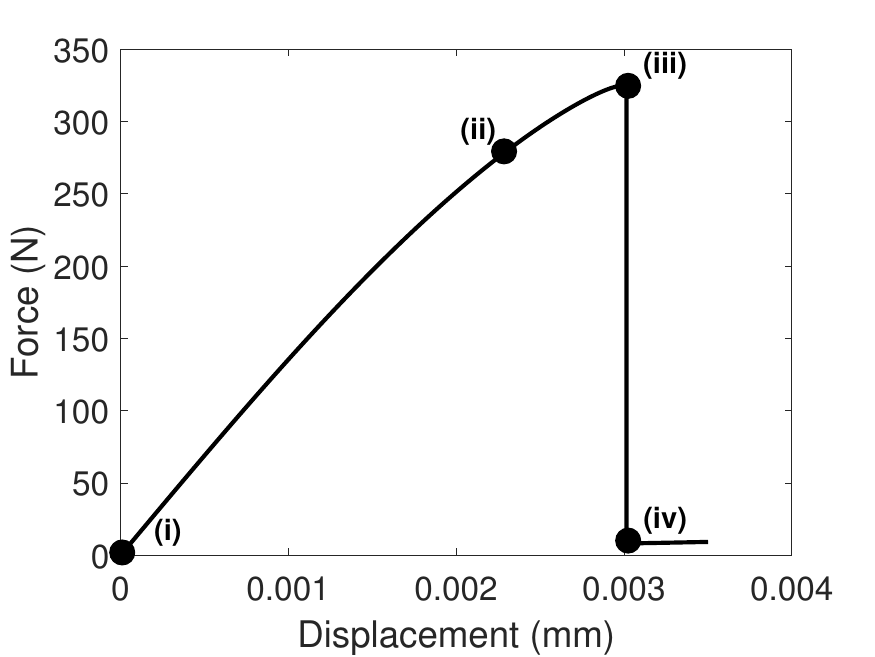}
\caption{Force-displacement results for the SENT using Abaqus user material subroutines (UMAT+UMATHT) developed in this work.}
\label{fig:SENT_F_D}
\end{figure}

\begin{figure}[htb]
\centering
\begin{tabular}{cc}
\includegraphics[width = .45\textwidth]{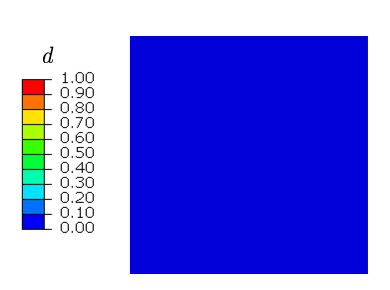} &
\includegraphics[width = .45\textwidth]{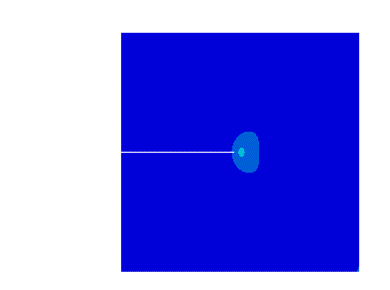} \\
\qquad \qquad (i) & \qquad \qquad (ii)
\end{tabular}
\begin{tabular}{cc}
\includegraphics[width = .45\textwidth]{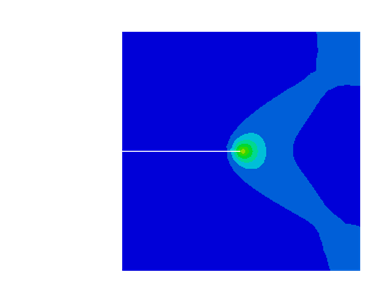} &
\includegraphics[width = .45\textwidth]{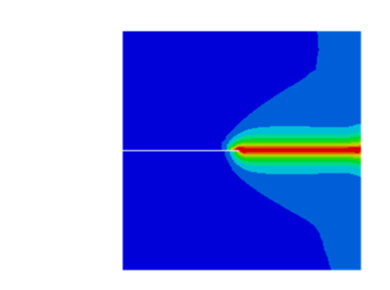} \\
\qquad \qquad (iii) & \qquad \qquad (iv)
\end{tabular}
\caption{Damage evolution for the SENT test using Abaqus user material subroutines (UMAT+UMATHT) developed in this work at different instants shown in Figure \ref{fig:SENT_F_D}.}
\label{fig:SENT_Damage_Evolution}
\end{figure}

It can be observed from Figure \ref{fig:SENT_F_D} that upon reaching a critical energy value, there is a sharp drop in the force-displacement response resulting in a brittle fracture, similar to what was reported in \citet{miehe2010thermodynamically}.
Given the geometry and applied boundary conditions, the fracture initiates near the left notch and propagates instantaneously through the specimen, leading to brittle failure, as expected and shown in Figure \ref{fig:SENT_Damage_Evolution}. This benchmark problem again validates our implementation of the gradient-damage theory for large deformation, which was developed in this work using Abaqus user material subroutines. It also demonstrates that the implementation, while formulated for large deformation kinematics as the most general case, remains applicable within the limits of small deformation kinematics.

\clearpage
\subsection{Single edge notched beam (SENB) test: large deformation rate-dependent plasticity}
\begin{figure}[htb]
\centering
	 \includegraphics[width = 0.7\textwidth]{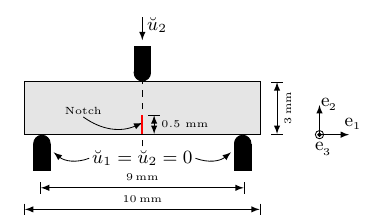}
\caption{Schematic showing the geometry and boundary conditions for the single edge notched beam.}
\label{fig:Schematic_SENB}
\end{figure}

\begin{figure}[htb]
\centering
\begin{tabular}{cc}
	 \includegraphics[width = 0.5\textwidth]{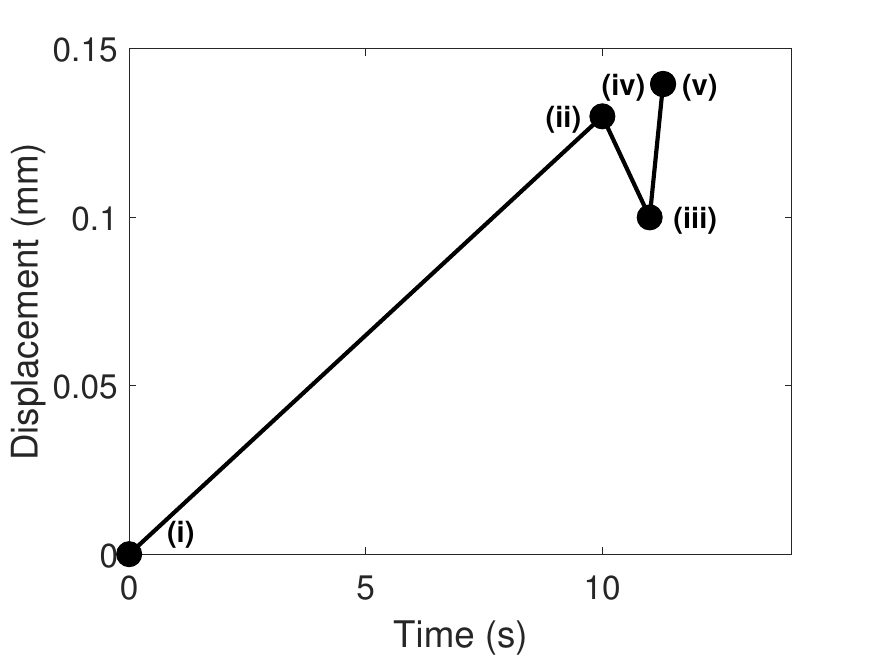} & \includegraphics[width = 0.5\textwidth]{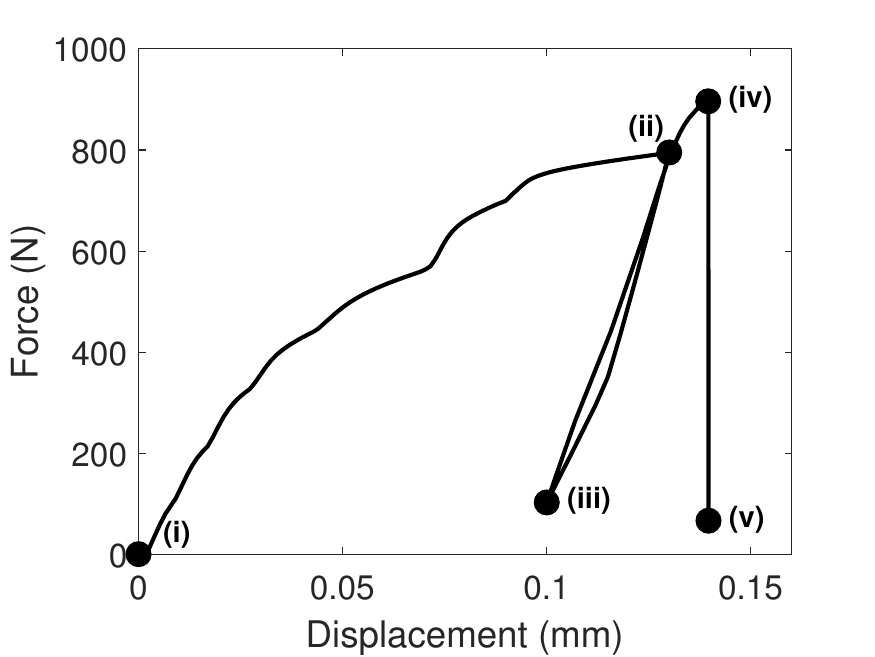} \\
     (a) & \qquad (b)
     \end{tabular}
\caption{SENB results using Abaqus user material subroutines (UMAT+UMATHT) developed in this work. (a) Prescribed displacement profile and (b) corresponding Force-displacement curve.}
\label{fig:SENB_F_D_T}
\end{figure}

In this last example, we examine the fracture response of a single edge notched beam of length $L_0 = 10\,$mm, and width $W_0 = 3\,$mm with a vertical notch of size $a = 0.5\,$mm, extending from the bottom-mid outer surface, as shown in Figure \ref{fig:Schematic_SENB}.
The bottom left and right supports are fixed, and displacement is applied to its top support, such that $u_2$ is prescribed as seen in Figure \ref{fig:SENB_F_D_T}a.
Frictional contact with a friction coefficient of $0.3$ is assumed to model the interaction between the support and the beam. All other surfaces are traction-free. In this example, the material constitutive response is taken to follow an isotropic rate-dependent elastic-plastic solid undergoing large deformation, with constitutive details and material parameters provided in Appendix \ref{Appendix:Constitutive_FeFp}.
We model this problem using a plane strain approximation. Specifically, the mesh consists of 17122 4-node linear displacement-temperature elements (CPE4T), with the elements along the notch path having a mesh size $l_e = 0.02\,$mm.

\begin{figure}[htb]
\centering
\begin{tabular}{cc}
$\HR$ at instant (i) & \qquad $\HR$ at instant (ii) \\
\includegraphics[width = .5\textwidth]{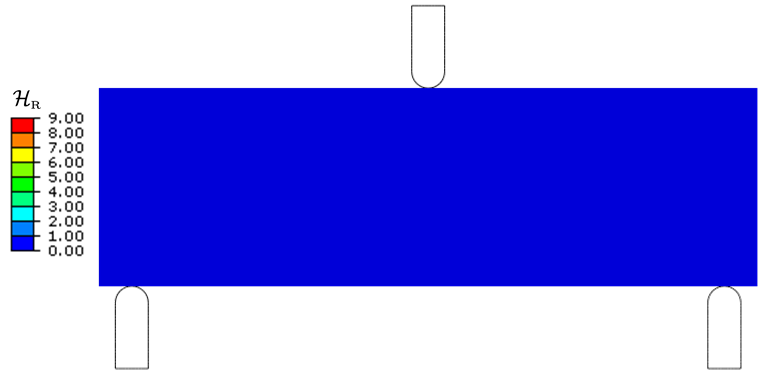} &
\includegraphics[width = .5\textwidth]{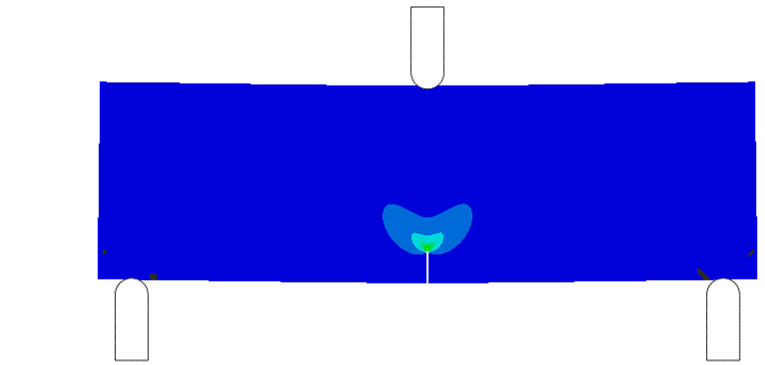}\\
(a) & \qquad (b) \\ \\
$\HR$ at instant (iv) & \qquad $\HR$ at instant (v) \\
\includegraphics[width = .5\textwidth]{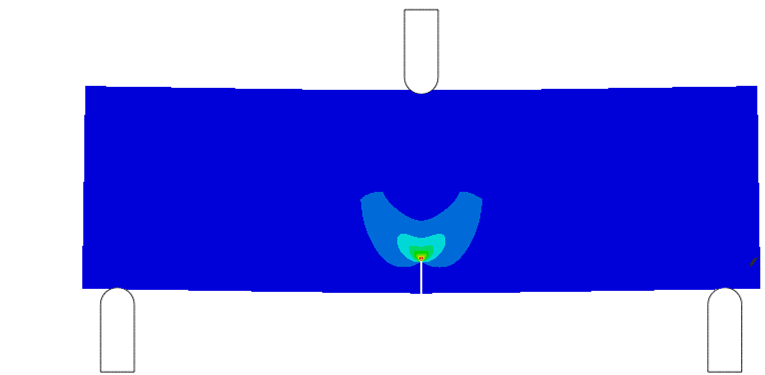} &
\includegraphics[width = .5\textwidth]{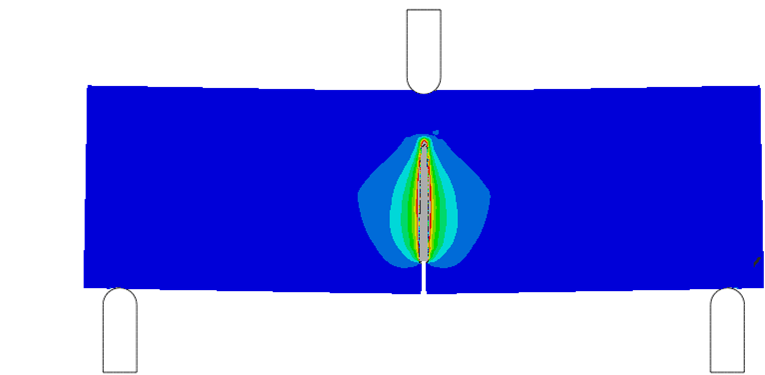}\\
(c) & \qquad (d)
\end{tabular}
\caption{History function ($\HR$) evolution for the SENB test using Abaqus user material subroutines (UMAT+UMATHT) developed in this work at different instants shown in Figure \ref{fig:SENB_F_D_T}. Note that the gray contour indicates regions where the values of $\HR$ exceed the limits shown in the legend.}
\label{fig:SENB_Hr_Evolution}
\end{figure}

\begin{figure}[htb]
\centering
\begin{tabular}{cc}
$d$ at instant (i) & \qquad $d$ at instant (ii) \\
\includegraphics[width = .5\textwidth]{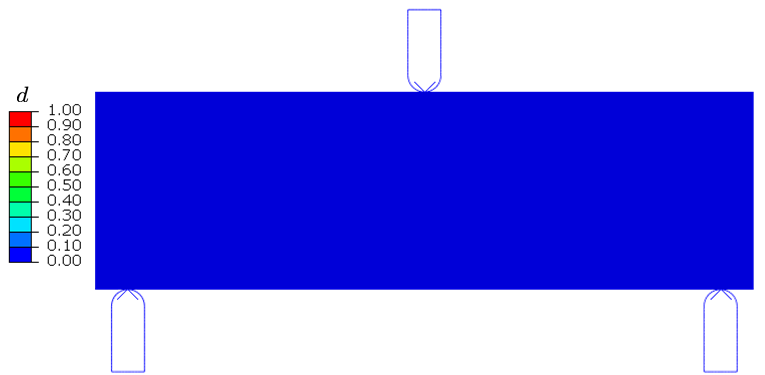} &
\includegraphics[width = .5\textwidth]{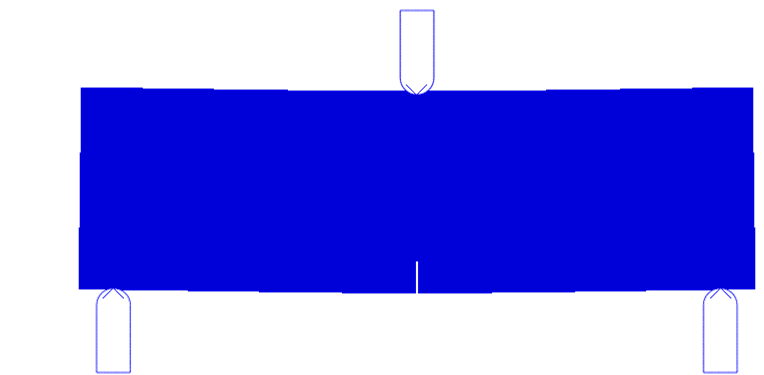}\\
(a) & \qquad (b) \\ \\
$d$ at instant (iv) & \qquad $d$ at instant (v) \\
\includegraphics[width = .5\textwidth]{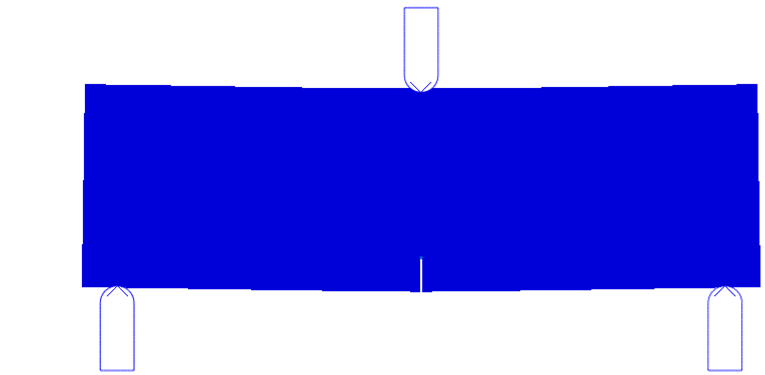} &
\includegraphics[width = .5\textwidth]{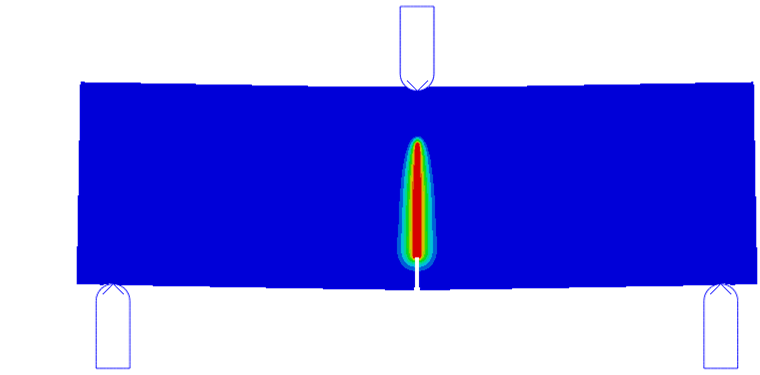}\\
(c) & \qquad (d)
\end{tabular}
\caption{Damage ($d$) evolution for the SENB test using Abaqus user material subroutines (UMAT+UMATHT) developed in this work at different instants shown in Figure \ref{fig:SENB_F_D_T}.}
\label{fig:SENB_Damage_Evolution}
\end{figure}

It can be observed from Figure \ref{fig:SENB_F_D_T} that upon the first loading cycle, i.e., loading from (i) to (ii) and unloading from (ii) to (iii), the material undergoes some permanent deformation that can be observed upon partial unloading and is attributed to plasticity. 
We note that in this first loading cycle, the energy captured by the history function $\HR$ and shown in Figure \ref{fig:SENB_Hr_Evolution}b remains below its critical value near the notch and thus no damage is observed to that region (see Figure \ref{fig:SENB_Damage_Evolution}b).
Moving onto the second loading cycle, i.e., loading from (iii) to (iv), the material is reloaded elastically until it reaches its prior maximum loading state, after which, it starts deforming plastically again. Upon further loading from (iv) to (v), the energy reaches and exceeds its critical value near the notch as can be observed in Figures \ref{fig:SENB_Hr_Evolution}c and d. Specifically, the damage nucleates near the notch on the bottom center and is quickly propagated through the specimen, causing a brittle failure captured by the force-displacement response and observed in Figures \ref{fig:SENB_Damage_Evolution}c and d.

This final benchmark problem further demonstrates the robustness and versatility of our implementation, highlighting its applicability across a wide range of materials, ranging from nonlinear elasticity to large deformation rate-dependent plasticity, making it valuable to engineers in various disciplines.

\clearpage

\section{Conclusion}\label{sec:Conclusion}
In this work, we have provided a pedagogic view of an appropriate Abaqus implementation of a gradient-damage theory for fracture in materials undergoing large deformation using Abaqus user material subroutines UMAT and UMATHT, which allows for accurately solving damage and failure problems without the need for writing complex UELs.

This was accomplished by carefully reviewing the analogy between the heat equation implemented in Abaqus and the PDE governing the gradient-damage theory in the context of \emph{large deformation}. The details of the implementation using the user material subroutines UMAT and UMATHT \citep{AbqStan} were provided and validated against a user element implementation from \citep{lee2023finite}. The UMAT and UMATHT implementation was further used to study classical benchmark problems from the literature, highlighting its robustness.

The Abaqus user material subroutines developed in this work and the input files are provided as supplemental materials to this paper. The implementation method and files provide researchers in the field with a simple yet robust tool to simulate and predict fracture and failure in solids, applicable to a broad class of materials through user-defined constitutive response.

\clearpage

\section*{CRediT authorship contribution statement}
\textbf{Keven Alkhoury}: Conceptualization, Methodology, Software, Formal analysis, Investigation, Writing - Original Draft. 
\textbf{Shawn A. Chester}: Conceptualization, Writing - Review \& Editing. 
\textbf{Vikas Srivastava}: Conceptualization, Supervision, Writing - Review \& Editing, Funding acquisition.

\section*{Acknowledgments}
KA acknowledges support from the Hibbitt Engineering Fellowship at Brown University.  VS acknowledges support from the US Office of Naval Research under grant number N00014-21-1-2815. SAC acknowledges partial support from the National Science Foundation, USA under grant number CMMI-1751520. The opinions, findings, conclusions, and recommendations expressed are those of the authors and do not necessarily reflect the views of the National Science Foundation or any US government agencies.

\section*{Code availability}
The code developed and used in this study will be made available upon publication.

\clearpage

\bibliographystyle{abbrvnat}
\bibliography{References}

\clearpage

\appendix
\section{Abaqus implementation using user material subroutines UMAT and UMATHT}\label{Appendix:Implementation}
\setcounter{table}{0}

In this appendix, we provide the details of the Abaqus implementation of the gradient-damage theory for large deformation fracture using user material subroutines UMAT and UMATHT.

Once again, to use the heat equation provided by Abaqus through the user subroutine UMAT, one needs to modify each of the terms in Table \ref{tab:comparison} (reproduced here for easier reading) as follows, with the caveat that the damage ``$d$'' is represented by temperature ``$\theta$'': 

\begin{table}[h]
    \centering
    \renewcommand{\arraystretch}{1.5} 
    \begin{tabular}{c|c|c|c}
       \textbf{Equation}  & \textbf{Transient term} & \textbf{Conduction term} & \textbf{Source term} \\ \hline
        \textbf{ \eqref{eq:HeatEquation_StrongForm}} & $\rho C \dot{\theta}$ & $\divx (\bfK \grad \theta)$ & $r$ \\ 
        \textbf{ \eqref{eqn:DamageEvolution_Spatial_RECAST}} & $J^{-1} \zetaR \ddot$ & $ \epsilonR l^{2} \divx (J^{-1} \bfB \grad d)$ & $2 (1 - d) J^{-1} \HR -  J^{-1} \epsilonR d$ \\
    \end{tabular}
    \caption{Term-by-term comparison of the heat equation \eqref{eq:HeatEquation_StrongForm} and the damage evolution equation \eqref{eqn:DamageEvolution_Spatial_RECAST}.}
    \label{tab:comparison_appendix_reproduced}
\end{table}

\begin{itemize}
    \item Starting with the transient term, one needs to impose the equality $\rho C  = J^{-1} \zetaR$, which requires access to the thermal constitutive behavior that can be achieved using the user subroutine ``UMATHT.''
    \item Similarly, one needs to make two changes to the conduction term such that:
\begin{enumerate}
    \item The thermal conductivity tensor $\bfK$ in Fourier's law must be replaced by $J^{-1} \bfB$.
    \item The conduction term must be multiplied by a pre-factor $\epsilonR l^{2}$.
\end{enumerate}

Both changes can also be done through the user subroutine ``UMATHT.''

\item Lastly, the source term needs to be $r = 2 (1 - d) J^{-1} \HR -  J^{-1} \epsilonR d$, which can be directly achieved through the ``RPL'' functionality in the user subroutine ``UMAT.''
\end{itemize}

Before proceeding, we present the UMATHT subroutine as provided in \citet{AbqStan} documentation for completeness.

\begin{lstlisting}[language=Fortran]
SUBROUTINE UMATHT(U,DUDT,DUDG,FLUX,DFDT,DFDG,
     1 STATEV,TEMP,DTEMP,DTEMDX,TIME,DTIME,PREDEF,DPRED,
     2 CMNAME,NTGRD,NSTATV,PROPS,NPROPS,COORDS,PNEWDT,
     3 NOEL,NPT,LAYER,KSPT,KSTEP,KINC)
C
      INCLUDE 'ABA_PARAM.INC'
C
      CHARACTER*80 CMNAME
      DIMENSION DUDG(NTGRD),FLUX(NTGRD),DFDT(NTGRD),
     1 DFDG(NTGRD,NTGRD),STATEV(NSTATV),DTEMDX(NTGRD),
     2 TIME(2),PREDEF(1),DPRED(1),PROPS(NPROPS),COORDS(3)
C
      COND = PROPS(1)
      SPECHT = PROPS(2)
C
      DUDT = SPECHT
      DU = DUDT*DTEMP
      U = U+DU
C
      DO I=1, NTGRD
         FLUX(I) = -COND*DTEMDX(I)
         DFDG(I,I) = -COND
      END DO
C
      RETURN
      END
\end{lstlisting}
Here, we note that ``SPECHT,'' first introduced in line 14 of the original code, represents the specific heat $C$ listed in Table \ref{tab:comparison_appendix_reproduced} such that ``SPECHT = $C$.'' Similarly, ``FLUX,'' appearing in line 21 of the original code, represents the heat flux computed using Fourier's law, following the constitutive relation ``FLUX=$-\bfK\grad\theta$'' also provided in Table \ref{tab:comparison_appendix_reproduced}.

In what follows, we refer to the original code provided by Abaqus documentation as ``original code'' and the modifications done in this work as ``modified code.'' 

\subsection{Transient term}\label{Appendix:Transient}
One could easily impose the equality $\rho C  = J^{-1} \zetaR$ through UMATHT by
\begin{enumerate}
    \item Prescribing a density $\rho = \zetaR$ and a specific heat $C = 1$ through the input file, so that $\rho C = \zetaR$.
    \item Passing $J=\det\bfF$ as a state variable (STATEV) from the UMAT subroutine and modifying line 16 in the original code such that 
\begin{itemize}
    \item original code: \begin{lstlisting}[language=Fortran, firstnumber=16]
      DUDT = SPECHT
    \end{lstlisting}
    \item modified code: \begin{lstlisting}[language=Fortran, firstnumber=16]
      DUDT = SPECHT/detF
    \end{lstlisting}
\end{itemize}
\end{enumerate}

\subsection{Conduction term}\label{Appendix:Conduction}
Two modifications are needed for the conduction term:
\begin{enumerate}
    \item To replace the thermal conductivity tensor $\bfK$ in Fourier's law by $J^{-1} \bfB$, we pass $\bfF$ as a state variable (STATEV) from the UMAT subroutine, which can then be used to calculate $ J^{-1} \bfB$.

    \item Moreover, we introduce a pre-factor $\epsilonR l^{2}$, referred to as ``AUX'', in the code and multiply the conduction term by it.
    \end{enumerate}

Both changes are reflected in the code below:
    
\begin{itemize}
    \item original code: \begin{lstlisting}[language=Fortran, firstnumber=20]
    DO I=1, NTGRD
        FLUX(I) = -COND*DTEMDX(I)
        DFDG(I,I) = -COND
    END DO
    \end{lstlisting}
    \item modified code: \begin{lstlisting}[language=Fortran, firstnumber=20]
    AUX = (erff*(lc**two))
    FLUX = zero
    DO J=1, NTGRD
        DO I=1, NTGRD
         FLUX(I) = FLUX(I) - (B(I,J)/detF)*DTEMDX(J)
        END DO 
    END DO
    FLUX = FLUX * AUX
C
    DFDG = ZERO
    DO I=1, NTGRD
        DO J=1, NTGRD
        DFDG(I,J) = DFDG(I,J) - (B(I,J)/detF)
        END DO
    END DO
    DFDG = AUX * DFDG

      \end{lstlisting}
\end{itemize}

\subsection{Source term}\label{Appendix:Source}
Lastly, the source term 
\begin{equation*}
r = 2 (1 - d) J^{-1} \HR -  J^{-1} \epsilonR d \end{equation*}
can be implemented through RPL in the UMAT subroutine with the details provided in the code (see supplemental materials). 

And, since DRPLDT ($\frac{\partial r}{\partial \theta} = \frac{\partial r}{\partial d}$), the variation of RPL with respect to temperature (damage) is needed for the Newton solver to ensure convergence in coupled temperature (damage) - displacement analyses, we also introduce 
\begin{equation}\label{eqn:drpldt}
  \frac{\partial r}{\partial d} = -2 \HR - \epsilonR\,.
\end{equation}

\begin{Remark}
    At first glance, \eqref{eqn:drpldt} appears to be missing $J^{-1}$. However, we encountered convergence issues when using an exact definition that includes $J^{-1}$, and thus, we have adopted \eqref{eqn:drpldt} throughout this work.
\end{Remark}

\clearpage

\section{Additional constitutive details}

In this appendix, we provide the additional constitutive details used in solving the boundary value problems.

\subsection{Compressible Neo-Hookean solid}\label{Appendix:Constitutive_Neo}

Following \citet{lee2023finite}, we take the free energy of the undamaged body such that
\begin{equation}\label{eqn:FreeEnergy_NeoCompressible}
    \hat{\psi}_\mat^{0} (\bfC) = \frac{G}{2} \left[ \Tr (\bfC) - 3  \right] + \frac{G}{\beta} \left[ (\det \bfF)^{-\beta} - 1  \right]\,,
\end{equation}
where $G$ represents the shear modulus,  $\beta \Def \frac{2 \nu_\text{poi}}{1 - 2 \nu_\text{poi}}$, with $\nu_\text{poi}$  
 representing Poisson's ratio. 
 Next, the Cauchy stress can be obtained using \eqref{eqn:LocalFreeEnergyGeneric}, \eqref{eqn:g(d)}, and \eqref{eqn:FreeEnergy_NeoCompressible} such that
 \begin{equation}    
\begin{split} \label{eq:CauchyNeohookean}
    \bfT & = 
    J^{-1} \bfF \big( 2 \frac{\partial \hat{\psi}_\mat^{\ast}}{\partial \bfC} )  {\bfF}^\trans \\ & = J^{-1} (1 - d)^2 \left[ G \big( \bfB - (\det F)^{-\beta} \id \big) \right] \,,
\end{split}
\end{equation}
with parameters tabulated in Table \ref{tab:BVP_MaterialParameters}.
\begin{table}[htb]
\centering
\begin{tabular}{cc|cc}
\multicolumn{2}{c}{Deformation} & \multicolumn{2}{c}{Damage} \\
\hline
Parameter & Value & Parameter & Value \\
\hline
$G$ (MPa) &  $5$ &  $\epsilonR$ (MPa) &  $240$ \\
$\nu_\text{poi}$ &  $0.45$ & $l$ (mm) &  $0.03$ \\ 
 &  & $\zetaR$ (MPa s) &  $0.001$ \\
\hline
\end{tabular}
\caption{Deformation and damage material parameters for a representative compressible Neo-Hookean solid.}
\label{tab:BVP_MaterialParameters}
\end{table}

\subsection{Linear elastic solid}\label{Appendix:Constitutive_LE}
We introduce the Green strain tensor 
\begin{equation}\label{eqn:GreenStrain}
    \bfE = \frac{1}{2} (\bfC - \id)\,,
\end{equation}
along with the displacement gradient
\begin{equation}
    \bfH = \bfF - \id \,,
\end{equation}
and use it to recast \eqref{eqn:GreenStrain} so that
\begin{equation}\label{eqn:GreenStrainDG}
    \bfE = \frac{1}{2} (\bfH + \bfH^\trans + \bfH \bfH^\trans)\,.
\end{equation}
And since we focus our attention on small deformations (or small strains), we drop the higher order terms in \eqref{eqn:GreenStrainDG} to obtain
\begin{equation}\label{eqn:smallStrain}
    \bfE = \frac{1}{2} (\bfH + \bfH^\trans) \,,
\end{equation}
and take the elastic free energy of the undamaged body such that
\begin{equation}\label{eqn:FreeEnergy_Elastic}
	 \hat{\psi}_\mat^{0} (\bfE)  = G \lvert {\bfE_{0}}\rvert ^{2} + \frac{1}{2} K(\Tr\bfE)^{2}\,,
\end{equation}
where
\begin{equation}\label{eqn:ShearModulus}
   G \Def \frac{E}{2 (1+\nu_\text{poi})}\,,
\end{equation}
is the shear modulus, and
\begin{equation}\label{eqn:BulkModulus}
     K \Def \frac{E}{3(1 - 2\nu_\text{poi})}\,,
\end{equation}
is the bulk modulus.
Next, the Cauchy stress can be obtained using \eqref{eqn:LocalFreeEnergyGeneric}, \eqref{eqn:g(d)}, and \eqref{eqn:FreeEnergy_Elastic} such that
 \begin{equation}    
 \label{eq:CauchyLinearElastic}
    \bfT  = \frac{\partial  \hat{\psi}_\mat^{\ast}}{\partial \bfE^{e}} = (1 - d)^2 \left[2G  \bfE_{0}^{e} + K (\Tr\bfE^{e}) \id  \right]\,,
\end{equation}
with parameters tabulated in Table \ref{tab:BVP_MaterialParameters_Elastic}.
\begin{table}[htb]
\centering
\begin{tabular}{cc|cc}
\multicolumn{2}{c}{Deformation} & \multicolumn{2}{c}{Damage} \\
\hline
Parameter & Value & Parameter & Value \\
\hline
$E$ (MPa) &  $ 210000$ &  $\epsilonR$ (MPa) &  $20.7$ \\
$\nu_\text{poi}$ &  $0.3$ & $l$ (mm) &  $0.03$ \\ 
 &  & $\zetaR$ (MPa s) &  $0.001$ \\
\hline
\end{tabular}
\caption{Deformation and damage material parameters for a representative linear elastic solid.}
\label{tab:BVP_MaterialParameters_Elastic}
\end{table}

\subsection{Large deformation rate-dependent plasticity}\label{Appendix:Constitutive_FeFp}
We assume in this example that damage nucleates only after the energy exceeds a critical value $\psi_\text{cr}$, and that damage initiation and propagation can only occur under tensile loading.
Accordingly, \eqref{eqn:HistoryFunction} is modified as follows 
\begin{equation}
    \HR(t) = 
    \begin{cases}
        \displaystyle\max_{s \in [0,t]}  \langle \hat{\psi}_{\mat}^{0} \big(\bfC(s)\big) - \psi_\text{cr}  \rangle & \text{if } J > 1 \\
        \HR(t-1) & \text{if } J \leq 1\,,
    \end{cases}
\end{equation}
where $\langle \bullet \rangle$ correspond to the Macaulay brackets and $\HR(t-1)$ to the history function in the prior step. Such a formulation allows damage to nucleate only after a critical energy threshold is reached, while preventing damage under compressive loading.\footnote{Other successful formulations for preventing damage under compressive loadings, such as the volumetric-deviatoric split  \citep{amor2009regularized} and the spectral decomposition \citet{miehe2010thermodynamically} have been reported in the literature.}

Next, following the previous work of Anand and co-workers \citep{Anand2009a,Srivastava2010}, we revisit \eqref{eqn:DefGrad}, and introduce the velocity gradient $\bfL$
\begin{equation}\label{eq:DefGrad}
  \bfF = \nabla \bfchi, \qquad \bfL = \grad  \bfv = \dot{\bfF}\bfF^{-1}\,.
\end{equation}
Moving forward, we adopt the standard multiplicative decomposition 
\begin{equation}\label{eq:FeFp}
  \bfF = \bfF^{e} \bfF^{p}, \qquad \text{with} \qquad \det\bfF^{e} > 0 \qquad \text{and} \qquad \det\bfF^{p} > 0\,,
\end{equation}
of the deformation gradient $\bfF$ into elastic $\bfF^{e}$ and plastic $\bfF^{p}$ parts.
With \eqref{eq:DefGrad} and \eqref{eq:FeFp}, we write the velocity gradient as 
\begin{equation}
\bfL = \dot{\bfF}\bfF^{-1} = {\bfL}^{e} + {\bfF}^{e} {\bfL}^{p} {{\bfF}^{e}}^{-1}
\end{equation}
where 
\begin{equation}
{\bfL}^{e} = \dot{\bfF}^{e}{{\bfF}^{e}}^{-1}, \qquad \text{and} \qquad {\bfL}^{p} = {\dot{\bfF}}^{p}{{\bfF}^{p}}^{-1}\,,
\end{equation}
are the elastic and plastic velocity gradients, respectively.
We also introduce the corresponding elastic and plastic stretching and spin tensors for completeness, such that 
\begin{equation}
\begin{split}
\bfD = \sym \bfL\,, \qquad  & \qquad \bfW = \skw \bfL\,, \\
\bfD^{e} = \sym \bfL^{e}\,, \qquad & \qquad \bfW^{e} = \skw \bfL^{e}\,, \\
\bfD^{p} = \sym \bfL^{p}\,, \qquad & \qquad \bfW^{p} = \skw \bfL^{p}\,, 
 \end{split}
\end{equation}
with $\bfL = \bfD + \bfW$, ${\bfL}^{e} = {\bfD}^{e} + {\bfW}^{e}$, and ${\bfL}^{p} = {\bfD}^{p} + {\bfW}^{p}$.
Further, since assuming an isotropic response, the plastic flow is taken to be irrotational such that 
\begin{equation}
\bfW^{p} = \bf0\,.
\end{equation}
As a consequence, $\bfL^{p} = \bfD^{p}$, and 
\begin{equation} \label{eq:FpDot}
\dot{\bfF}^{p} = \bfD^{p} \bfF^{p}\,.
\end{equation}
And lastly, plastic flow is assumed to be incompressible such that
\begin{equation}
J^{p} = 1,
\end{equation}
and as a result
\begin{equation}
\text{tr} \bfL^{p} = \text{tr} \bfD^{p} = 0\,,
\end{equation}
and therefore 
\begin{equation}
J^{e} = J\,.
\end{equation}
We pick the standard form of the free energy \emph{per unit intermediate volume} 
\begin{equation} \label{eqn:FreeEnergy_Elastic_LD_Inter}
 \psi^{e} (\bfELN) = G\lvert \bfELN_{0}^{e}\rvert ^{2} + \frac{1}{2} K (\text{tr}\bfELN^{e})^{2}\,, 
\end{equation}
where $G$ and $K$ are the shear and bulk moduli, given in \eqref{eqn:ShearModulus} and \eqref{eqn:BulkModulus}, respectively, and $\bfELN_{0}^{e}$ is the deviatoric part of the logarithmic elastic strain $\bfELN^{e}=\ln\bfU^e$.

Following \citet{narayan2021fracture}, we take the undamaged energy \emph{per unit reference volume} such that \begin{equation}\label{eqn:FreeEnergy_Elastic_LD_Ref}
\hat{\psi}_\mat^{0} (\bfELN) = J^{p} \psi^{e} (\bfELN) =  \psi^{e} (\bfELN) \,.
\end{equation}
Next, the corresponding Mandel stress can be obtained using \eqref{eqn:LocalFreeEnergyGeneric}, \eqref{eqn:g(d)}, \eqref{eqn:FreeEnergy_Elastic_LD_Inter}, and \eqref{eqn:FreeEnergy_Elastic_LD_Ref} such that
\begin{equation}
\bfM^{e} = \frac{\partial \hat{\psi}_\mat^{\ast}}{\partial \bfELN^{e}} = (1 - d)^2 \left[2G \bfELN_{0}^{e} + K (\text{tr\bfELN}^{e}) \id \right] \,.
\end{equation}
We focus our attention on isotropic materials, hence, the corresponding Cauchy stress is related to the Mandel stress by
\begin{equation}
\bfT \Def J^{-1} \bfR^{e} \bfM^{e} \bfR^{e\trans}\,.
\end{equation}
Now, keeping \eqref{eq:FeFp} and \eqref{eq:FpDot} in mind, we write the flow rule in the shear-driven form
\begin{equation}
\bfD^{p} = \sqrt{\frac{1}{2}} \nup \bfN^{p}\,,
\end{equation}
where $\nu^{p}$ and $\bfN^{p}$ are the equivalent plastic shear strain rate and the direction of plastic flow, respectively, and are given by
\begin{equation}\label{eq:EqRate}
\nu^{p} = \nu_{0} \Big(\frac{\bar{\tau}}{(1 - d)^2 S}\Big)^{1/m}\,,
\end{equation} 
and
\begin{equation}\label{eq:DirectionFlow}
\bfN^{p} = \frac{\bfM_{0}^{e}}{\sqrt{2}\bar{\tau}}\,,
\end{equation}
where $\bfM_{0}^{e}$ is the deviator of the Mandel stress. Moreover,
\begin{equation}
\bar{\tau} \Def \sqrt{\frac{1}{2} \bfM_{0}^{e} \tendot \bfM_{0}^{e}}\,,
\end{equation} 
is defined as the equivalent shear stress, and $S$ is the deformation resistance. The evolution equation of the deformation resistance is taken to follow 
\begin{equation}
    \dot{S} = h (S_\text{sat} - S) \nup\,, \quad S \left( \bfx, t=0 \right) = S_{0}\,,
\end{equation}
where $h$, $S_\text{sat}$, and $S_{0}$ represent the hardening slope, the saturation level of the deformation resistance, and the initial yield strength, respectively.
Lastly, material parameters are tabulated in Table \ref{tab:BVP_MaterialParameters_ElasticPlastic}.
\begin{table}[htb]
\centering
\begin{tabular}{cc|cc}
\multicolumn{2}{c}{Deformation} & \multicolumn{2}{c}{Damage} \\
\hline
Parameter & Value & Parameter & Value \\
\hline
$E$ (MPa) &  $ 210000$ &  $\epsilonR$ (MPa) &  $5$ \\
$\nu_\text{poi}$ &  $0.3$ & $l$ (mm) &  $0.1$ \\ 
$\nu_{0}$ (1/s) & 10$^{-4}$ & $\zetaR$ (MPa s) &  $0.001$ \\
$m$ & 0.05 & $\psi_\text{cr}$ (MPa) & $9$ \\
$h$ & 5 & & \\
$S_\text{sat}$ (MPa) & $450$ & & \\
$S_\text{0}$ (MPa) & $300$ & & \\
\hline
\end{tabular}
\caption{Deformation and damage material parameters for a representative rate-dependent plastic solid undergoing large deformation.}
\label{tab:BVP_MaterialParameters_ElasticPlastic}
\end{table}

\end{document}